\documentclass[12pt]{article}
\usepackage{amsmath}
\usepackage{amssymb}
\usepackage{graphicx}
\usepackage{axodraw}
\usepackage{epsfig}
\usepackage{subfigure}
\usepackage{fancybox}
\usepackage{url}
\usepackage[small]{caption}
\begin{document}
\baselineskip 0.6cm

\def\simgt{\mathrel{\lower2.5pt\vbox{\lineskip=0pt\baselineskip=0pt
           \hbox{$>$}\hbox{$\sim$}}}}
\def\simlt{\mathrel{\lower2.5pt\vbox{\lineskip=0pt\baselineskip=0pt
           \hbox{$<$}\hbox{$\sim$}}}}
\def\simprop{\mathrel{\lower3.0pt\vbox{\lineskip=1.0pt\baselineskip=0pt
             \hbox{$\propto$}\hbox{$\sim$}}}}
\def\bra#1{\left< #1 \right|}
\def\ket#1{\left| #1 \right>}
\def\inner#1#2{\left< #1 | #2 \right>}
\def\observer{\begin{picture}(15,15)
  \CArc(7.5,7.5)(2.5,0,360)
  \put(5.2,4.9){$\cdot$} \Line(5.0,7.8)(4.6,6.6) \Line(6.6,6.2)(5.5,6.0)
  \Line(7.5,4.5)(3,1.5) \Line(7.5,4.5)(12,1.5) \Line(7.5,4.5)(7.5,0.5)
  \Line(7.5,0.5)(5,-5) \Line(7.5,0.5)(10,-5)
\end{picture}}
\def\brain{\begin{picture}(21,15)
  \Line(-3,9)(17,9) \CArc(17,7)(2,0,90) \Line(19,7)(19,-1.2)
  \CArc(-3,5)(4,0,90) \Line(1,5)(1,-1.2)
  \CArc(3,-1.2)(2,180,270) \Line(3,-3.2)(17,-3.2) \CArc(17,-1.2)(2,270,360)
  \Vertex(3.2,5.1){1} \put(5.7,4.0){\tiny ${\bf x}_{{\rm B\!H}}$}
  \Photon(4.4,3.7)(8.5,-0.4){0.4}{3}
  \Line(8.7,-0.8)(8.5,1.4) \Line(8.9,-0.6)(6.7,-0.4)
  \put(9.7,-1.7){\tiny $n$}
\end{picture}}
\def\chairup{\begin{picture}(15,15)
  \Line(4,10)(11,10) \Line(4,4)(11,4) \Line(4,4)(4,10) \Line(11,4)(11,10)
  \Line(4,4)(2.5,1) \Line(11,4)(12.5,1) \Line(2.5,1)(12.5,1)
  \Line(3,1)(3,-5) \Line(12,1)(12,-5)
  \Line(4.5,1)(4.5,-2.5) \Line(10.5,1)(10.5,-2.5)
\end{picture}}
\def\chairdown{\begin{picture}(15,15)
  \Line(4,-5)(11,-5) \Line(4,1)(11,1) \Line(4,1)(4,-5) \Line(11,1)(11,-5)
  \Line(4,1)(2.5,4) \Line(11,1)(12.5,4) \Line(2.5,4)(12.5,4)
  \Line(3,4)(3,10) \Line(12,4)(12,10)
  \Line(4.5,4)(4.5,7.5) \Line(10.5,4)(10.5,7.5)
\end{picture}}
\def\brainup{\begin{picture}(15,15)
  \Line(-3,9)(11,9) \CArc(11,7)(2,0,90) \Line(13,7)(13,-0.2)
  \CArc(-3,5)(4,0,90) \Line(1,5)(1,-0.2)
  \CArc(3,-0.2)(2,180,270) \Line(3,-2.2)(11,-2.2) \CArc(11,-0.2)(2,270,360)
  \Line(5,7)(8.5,7) \Line(5,4)(8.5,4)
  \Line(5,4)(5,7) \Line(8.5,4)(8.5,7)
  \Line(5,4)(4.25,2.5) \Line(8.5,4)(9.25,2.5) \Line(4.25,2.5)(9.25,2.5)
  \Line(4.5,2.5)(4.5,-0.5) \Line(9,2.5)(9,-0.5)
  \Line(5.25,2.5)(5.25,0.75) \Line(8.25,2.5)(8.25,0.75)
\end{picture}}
\def\braindown{\begin{picture}(15,15)
  \Line(-3,9)(11,9) \CArc(11,7)(2,0,90) \Line(13,7)(13,-0.2)
  \CArc(-3,5)(4,0,90) \Line(1,5)(1,-0.2)
  \CArc(3,-0.2)(2,180,270) \Line(3,-2.2)(11,-2.2) \CArc(11,-0.2)(2,270,360)
  \Line(5,-0.5)(8.5,-0.5) \Line(5,2.5)(8.5,2.5)
  \Line(5,2.5)(5,-0.5) \Line(8.5,2.5)(8.5,-0.5)
  \Line(5,2.5)(4.25,4) \Line(8.5,2.5)(9.25,4) \Line(4.25,4)(9.25,4)
  \Line(4.5,4)(4.5,7) \Line(9,4)(9,7)
  \Line(5.25,4)(5.25,5.75) \Line(8.25,4)(8.25,5.75)
\end{picture}}

\begin{titlepage}

\begin{flushright}
MIT-CTP-4405 \\
UCB-PTH-12/17 \\
\end{flushright}

\vskip 1.0cm

\begin{center}
{\Large \bf Black Holes, Information, and Hilbert Space\\
for Quantum Gravity}

\vskip 0.7cm

{\large Yasunori Nomura$^{a,b}$, Jaime Varela$^{a,b}$, and Sean J. Weinberg$^b$}

\vskip 0.4cm

$^a$ {\it Center for Theoretical Physics, Laboratory for Nuclear Science, 
     and Department of Physics, \\
     Massachusetts Institute of Technology, Cambridge, MA 02139, USA} \\

\vskip 0.2cm

$^b$ {\it Berkeley Center for Theoretical Physics, Department of Physics, \\
     and Theoretical Physics Group, Lawrence Berkeley National Laboratory, \\
     University of California, Berkeley, CA 94720, USA} \\

\vskip 0.8cm

\abstract{A coarse-grained description for the formation and evaporation 
 of a black hole is given within the framework of a unitary theory 
 of quantum gravity preserving locality, without dropping the 
 information that manifests as macroscopic properties of the state 
 at late times.  The resulting picture depends strongly on the 
 reference frame one chooses to describe the process.  In one description 
 based on a reference frame in which the reference point stays outside 
 the black hole horizon for sufficiently long time, a late black hole 
 state becomes a superposition of black holes in different locations 
 and with different spins, even if the back hole is formed from collapsing 
 matter that had a well-defined classical configuration with no angular 
 momentum.  The information about the initial state is partly encoded in 
 relative coefficients---especially phases---of the terms representing 
 macroscopically different geometries.  In another description in which 
 the reference point enters into the black hole horizon at late times, 
 an $S$-matrix description in the asymptotically Minkowski spacetime is 
 not applicable, but it sill allows for an ``$S$-matrix'' description 
 in the full quantum gravitational Hilbert space including singularity 
 states.  Relations between different descriptions are given by unitary 
 transformations acting on the full Hilbert space, and they in general 
 involve superpositions of ``distant'' and ``infalling'' descriptions. 
 Despite the intrinsically quantum mechanical nature of the black hole 
 state, measurements performed by a classical physical observer are 
 consistent with those implied by general relativity.  In particular, 
 the recently-considered firewall phenomenon can occur only for an 
 exponentially fine-tuned (and intrinsically quantum mechanical) initial 
 state, analogous to an entropy decreasing process in a system with 
 large degrees of freedom.}

\end{center}
\end{titlepage}

\section{Introduction}
\label{sec:intro}

Since its discovery~\cite{Hawking:1974sw}, the process of black 
hole formation and evaporation has contributed tremendously to our 
understanding of quantum aspects of gravity.  Building on earlier ideas, 
in particular the holographic principle~\cite{'tHooft:1993gx,Bousso:1999xy} 
and complementarity~\cite{Susskind:1993if}, one of the authors (Y.N.) has 
recently proposed an explicit framework for formulating quantum gravity 
in a way that is consistent with locality at length scales larger than 
the Planck (or string) length~\cite{Nomura:2011rb}. (Essentially the same 
idea had been used earlier to describe the eternally inflating multiverse 
in Ref.~\cite{Nomura:2011dt}.)  This allows us to describe a system with 
gravity in such a way that the picture based on conventional quantum 
mechanics, including the emergence of classical worlds due to amplification 
in position space~\cite{q-Darwinism,Nomura:2011rb}, persists without 
any major modification.

In the framework given in Ref.~\cite{Nomura:2011rb}, quantum states 
allowing for spacetime interpretation represent only the limited spacetime 
region in and on the apparent horizon {\it as viewed from a fixed (freely 
falling) reference frame}.  Complementarity, as well as the observer 
dependence of cosmic horizons, can then be understood---and thus precisely 
formulated---as a reference frame change represented by a unitary 
transformation acting on the full quantum gravitational Hilbert space, 
which takes the form
\begin{equation}
  {\cal H}_{\rm QG} = \left( \bigoplus_{\cal M} {\cal H}_{\cal M} \right) 
    \oplus {\cal H}_{\rm sing}.
\label{eq:H_QG}
\end{equation}
Here, ${\cal H}_{\cal M}$ is the Hilbert subspace containing states on 
a fixed semi-classical geometry ${\cal M}$ (or more precisely, a set of 
semi-classical geometries ${\cal M} = \{ {\cal M}_i \}$ having the 
same horizon $\partial {\cal M}$), while ${\cal H}_{\rm sing}$ is that 
containing ``intrinsically quantum mechanical'' states associated with 
spacetime singularities.%
\footnote{Here, the geometry means that of a codimension-one hypersurface 
 which the quantum states represent, and {\it not} that of spacetime.}
In its minimal implementation---which we assume throughout---the framework 
of Ref.~\cite{Nomura:2011rb} says that a state that is an element of one 
of the ${\cal H}_{\cal M}$'s represents a physical configuration on the 
past light cone of the origin, $p$, of the reference frame.  We call the 
structure of the Hilbert space in Eq.~(\ref{eq:H_QG}) with this particular 
implementation the {\it covariant Hilbert space for quantum gravity}.

The purpose of this paper is to develop a complete picture of black hole 
formation and evaporation in this framework, based on Eq.~(\ref{eq:H_QG}). 
Our picture says that:
\begin{itemize}
\item
The evolution of the full quantum state is unitary.
\item
The state, however, is in general a superposition of macroscopically 
different worlds.  In particular, the final state of black hole evaporation 
is a superposition of macroscopically distinguishable terms, even if 
the initial state forming the black hole is a classical object having 
a well-defined macroscopic configuration.  The information of the initial 
state is encoded partly in relative coefficients, especially in phases, 
among these macroscopically different terms.
\item
No physical observer can recover the initial state forming the black hole 
by observing final Hawking radiation quanta.  This is true even if the 
measurement is performed with arbitrarily high precision using an arbitrary 
(in general quantum) measuring device.
\item
Observations each physical observer makes are well described by the 
semi-classical picture in the regime it is supposed to be applicable, 
unless the observer (or measuring device) is in an exponentially rare 
quantum state in the corresponding Hilbert space.
\end{itemize}
We note that while some of the considerations here are indeed specific 
to the present framework, some are more general and apply to other theories 
of gravity as well, especially to the ones in which the formation and 
evaporation of a black hole is described as a unitary quantum mechanical 
process.

There are several key ingredients to understand the features described 
above, which we now highlight:
\begin{itemize}
\item
Quantum mechanics has a ``dichotomic'' nature about locality:\ while 
the {\it dynamics}, encoded in the time evolution operator, is local, 
a {\it state} is generically non-local, as is clearly demonstrated in 
the Einstein-Podolsky-Rosen experiment.  In particular, this allows 
for a state to be a superposition of terms representing macroscopically 
different spacetime geometries.
\item
The framework of Ref.~\cite{Nomura:2011rb} says that general covariance 
implies the quantum states must be defined in a fixed (local Lorentz) 
reference frame; moreover, to preserve locality of the dynamics ({\it not} 
of states), these states represent spacetime regions only in and on the 
stretched/apparent horizon as viewed from the reference frame.%
\footnote{Note that the apparent horizon, defined as a surface on which 
 the expansion of the past-directed light rays emitted from $p$ turns 
 from positive to negative, is in general {\it not} the same as the 
 event horizon of the black hole.  For example, there is no apparent, 
 or stretched, horizon associated with the black hole when the system 
 is described in an infalling reference frame.}
This implies, in particular, that once a reference frame is fixed, 
the {\it location} of a black hole (with respect to the reference frame) 
is a physically meaningful quantity, even if there is no other object.
\item
The location of a black hole is highly uncertain after long 
time~\cite{Page:1979tc}.  In particular, at a timescale of evaporation 
$\sim M(0)^3$, where $M(0)$ is the initial black hole mass, the uncertainty 
of the location is of order $M(0)^2$, which is much larger than the 
Schwarzschild radius of the initial black hole, $R_S = 2M(0)$.  (This 
is also the timescale in which the black hole loses more than a half 
of its initial Bekenstein-Hawking entropy in the form of Hawking 
radiation~\cite{Page:1993wv}.)  This implies that a state of a sufficiently 
old black hole becomes a superposition of terms representing macroscopically 
distinguishable worlds~\cite{Nomura:2012sw}.
\end{itemize}
As discussed in Refs.~\cite{Nomura:2011dt,Nomura:2012zb}, these ingredients, 
especially the first two, are also important to understand the eternally 
inflating multiverse (or quantum many universes) and to give well-defined 
probabilistic predictions in such a cosmology.

In this paper, we will first discuss the picture presented above in 
the case that the evolution of a black hole is described in a distant 
reference frame, i.e.\ a freely falling frame whose origin $p$ stays 
outside the black hole horizon for all time.  We will, however, also 
discuss in detail what happens if we describe the system using an 
infalling reference frame, i.e.\ a frame in which $p$ enters into the 
black hole horizon at late times.  Following Ref.~\cite{Nomura:2011rb}, 
we treat this problem by performing a unitary transformation on the state 
representing the black hole evolution as viewed from a distant reference 
frame.  We find that, because of the uncertainty of the black hole 
location, the resulting state is in general {\it a superposition of} 
infalling and distant descriptions of the black hole, and this effect 
is particularly significant when we try to describe the interior of 
an old black hole.

The issue of information in black hole evaporation has a long 
history of extensive research, with earlier proposals including 
information loss~\cite{Hawking:1976ra,Wald:1980nm}, 
remnants~\cite{Aharonov:1987tp,Giddings:1992hh}, 
and baby universes~\cite{Dyson:1976}; see, e.g., 
Refs.~\cite{Preskill:1992tc,Mathur:2009hf} for reviews. 
Our picture here is that the evolution of a black hole 
is unitary~\cite{'tHooft:1990fr}, as in the cases of the 
complementarity~\cite{Susskind:1993if} and fuzzball~\cite{Lunin:2001jy} 
pictures, with a macroscopically non-local nature of a black hole final 
state explicitly taken into account.  This non-locality is especially 
important for the explicit realization of complementarity as a 
unitary reference frame change in the quantum gravitational Hilbert 
space~\cite{Nomura:2011rb}.  We note that some authors have discussed 
non-locality at a macroscopic level~\cite{Giddings:1992hh,Giddings:2009ae} 
or an intrinsically quantum nature of black hole 
states~\cite{Brustein:2012jn} but in ways different 
from the ones considered here.  In particular, these proposals 
lead either to non-locality of the dynamics (not only states), 
large deviations from the semi-classical picture experienced by 
a macroscopic physical observer, or intrinsic quantum effects 
confined only to the microscopic domain, none of which applies 
to the picture presented here.

The organization of this paper is as follows.  In Section~\ref{sec:BH}, 
we provide a detailed description of the formation and evaporation of 
a black hole as viewed from a distant reference frame.  We elucidate 
the meaning of information in this context, and find that it is partly 
in relative coefficients of terms representing different macroscopic 
configurations of Hawking quanta and/or geometries.  We also discuss 
what a physical observer, who should be a part of the description, 
will measure and if he/she can reconstruct the initial state based 
on his/her measurements.  In Section~\ref{sec:compl}, we consider 
descriptions of the system in different reference frames.  In particular, 
we discuss how the spacetime inside the black hole horizon appears 
in these descriptions and how these descriptions are related to the 
distant description given in Section~\ref{sec:BH}.  We also elaborate 
on the analysis of Ref.~\cite{Nomura:2012sw}, arguing that the firewall 
paradox recently pointed out in Ref.~\cite{Almheiri:2012rt} does not 
exist (although the firewall ``phenomenon'' can occur if the initial 
state is exponentially fine-tuned).  In Section~\ref{sec:concl} we give 
our final discussion and conclusions.  In the appendix, we provide an 
analysis of the ``spontaneous spin-up'' phenomenon, which we find to 
occur for a general Schwarzschild (or very slowly rotating) black hole. 
This effect makes a Schwarzschild black hole evolve into a superposition 
of Kerr black holes with distinct angular momenta.

Throughout the paper, we limit our discussions to the case of four 
spacetime dimensions to avoid unnecessary complications of various 
expressions; but the extension to other dimensions is straightforward. 
We also take the unit in which the Planck scale, $G_N^{-1/2} \simeq 
1.22 \times 10^{19}~{\rm GeV}$, is set to unity, so all the quantities 
appearing can be regarded as dimensionless.

\section{Black Holes and Unitarity---A Distant View}
\label{sec:BH}

In this section we discuss how unitarity of quantum mechanical evolution 
is preserved in the process in which a black hole forms and evaporates, 
as viewed from a distant reference frame.  We clarify the meaning of the 
information in this context, and argue that it (partly) lies in relative 
coefficients---especially phases---of terms representing {\it macroscopically 
distinct} configurations in a full quantum state.  We also elucidate 
the fact that a physical observer can never extract complete (quantum) 
information of the initial state forming the black hole; i.e., observing 
final-state Hawking radiation does not allow for him/her to infer the 
initial state, despite the fact that the evolution of the entire quantum 
state is fully unitary.

\subsection{Information paradox---what is the information?}
\label{subsec:info_para}

In his famous 1976 paper, Hawking argued, based on semi-classical 
considerations, that a black hole loses information~\cite{Hawking:1976ra}. 
Consider two objects having the same energy-momentum, represented by 
pure quantum states $\ket{A}$ and $\ket{B}$, which later collapse into 
black holes with the same mass $M(0)$.  According to the semi-classical 
picture, the evolutions of the two states after forming the black holes 
are identical, leading to the same mixed state $\rho_H$, obtained by 
integrating the thermal Hawking radiation states:
\begin{equation}
\begin{array}{lll}
  \ket{A} &\rightarrow& \rho_H, \\
  \ket{B} &\rightarrow& \rho_H.
\end{array}
\label{eq:non-unitary}
\end{equation}
This phenomenon is referred to as the information loss in black holes.

What is the problem of this picture?  The problem is that since the 
final states are identical, we cannot recover the initial state of the 
evolution just by knowing the final state, even in principle.  This 
contradicts unitarity of quantum mechanical evolution, which says that 
time evolution of a state is reversible, i.e.\ we can always recover 
the initial state if we know the final state exactly by applying the 
inverse time evolution operator $e^{+iHt}$.

Based on various circumstantial evidence, especially AdS/CFT 
duality~\cite{Maldacena:1997re}, we now do not think the above picture 
is correct.  We think that the final states obtained from different initial 
states differ, and a state obtained by evolving any pure state is always 
pure even if the evolution involves formation and evaporation of a black 
hole.  Namely, instead of Eq.~(\ref{eq:non-unitary}), we have
\begin{equation}
\begin{array}{lll}
  \ket{A} &\rightarrow& \ket{\psi_A}, \\
  \ket{B} &\rightarrow& \ket{\psi_B},
\end{array}
\label{eq:unitary}
\end{equation}
where $\ket{\psi_A} \neq \ket{\psi_B}$ iff $\ket{A} \neq \ket{B}$.  In 
this picture, quantum states representing black holes formed by different 
initial states are different, even if they have the same mass.  (The 
dimension of the Hilbert space corresponding to a classical black hole 
of a fixed mass $M$ is $\exp({\cal A}_{\rm BH}/4)$ according to the 
Bekenstein-Hawking entropy, where ${\cal A}_{\rm BH} = 16\pi M^2$ is 
the area of the black hole horizon.)  These states then evolve into 
different final states $\ket{\psi_A}$ and $\ket{\psi_B}$, representing 
states for emitted Hawking radiation quanta.

A question is in what form the information is encoded in the final state. 
On one hand, possible final states of evaporation of a black hole must 
have a sufficient variety to encode complete information about the 
initial state forming the black hole.  This requires that the dimension 
of the Hilbert space corresponding to these states must be of order 
$\exp({\cal A}_{\rm BH}(0)/4)$, where ${\cal A}_{\rm BH}(0) = 16\pi 
M(0)^2$ is the area of the black hole horizon right after the formation. 
On the other hand, Hawking radiation quanta emitted from the black hole 
must have the thermal spectrum (with temperature $T_H = 1/8\pi M$ when 
the black hole mass is $M$) in the regime where the semi-classical 
analysis is valid, $M \gg 1$.  It is not clear how the state actually 
realizes these two features~\cite{Page:1993up}, although the generalized 
second law of thermodynamics guarantees that it can be done.  Below, we 
argue that a part of the information that is necessary to recover the 
initial state is contained in relative coefficients of terms representing 
different macroscopic worlds, even if the initial state has a well-defined 
classical configuration.

Our analysis does not prove unitarity of the black hole 
formation/evaporation process, or address the question of how the complete 
information of the initial state is encoded in the emitted Hawking 
quanta at the microscopic level.  Rather, we assume that unitarity is 
preserved at the microscopic level, and study manifestations of this 
assumption when we describe the process at a semi-classical level.  This 
will provide implications on how such a description must be constructed. 
For example, in order to preserve all the information in the initial 
state, the description must not be given on a fixed black hole background 
in an intermediate stage of the evaporation, since it would correspond 
to ignoring a part of the information contained in the full quantum state 
manifested as macroscopic properties of the remaining black hole.  Note 
that we do not claim that these macroscopic properties contain independent 
information beyond what is in the emitted Hawking quanta---the two are 
certainly correlated by energy-momentum conservation.  The analysis 
presented here also has implications on the complementarity picture, 
which will be discussed in Section~\ref{sec:compl}.

\subsection{Where is the information in the black hole state?}
\label{subsec:where_info}

\begin{figure}[t]
\begin{center}
  \includegraphics[width=12cm]{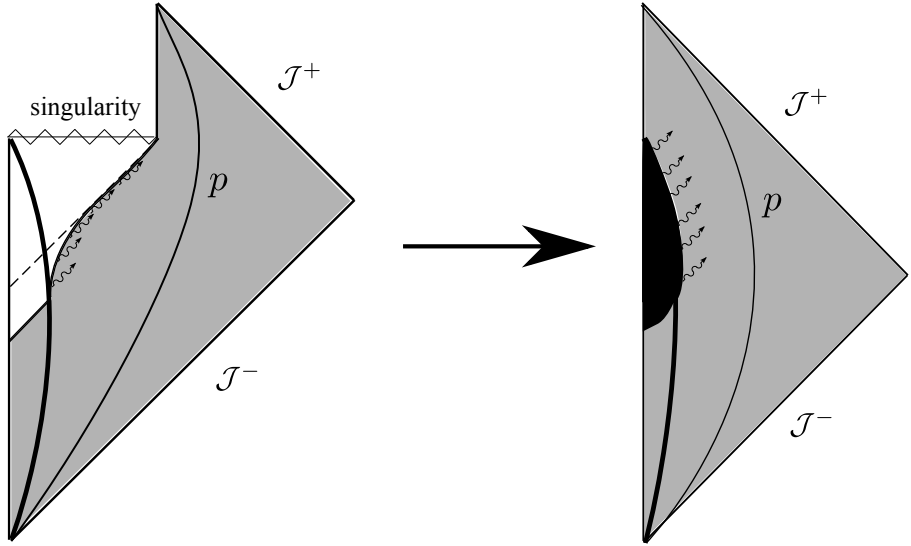}
\caption{The Penrose diagram representing a black hole formed from a 
 collapsing shell of matter (represented by the thick solid curve) which 
 then evaporates.  The left panel shows the standard ``global spacetime'' 
 picture, in which Hawking radiation (denoted by wavy arrows) comes 
 from the stretched horizon.  To obtain a consistent quantum mechanical 
 description, we must fix a reference frame (freely falling frame) and 
 then describe the system from that viewpoint~\cite{Nomura:2011rb}. 
 Quantum states then corresponds to physical configurations in the past 
 light cone of the origin $p$ of that reference frame.  Here we choose 
 a ``distant'' reference frame; the trajectory of its origin $p$ is 
 depicted by a thin solid curve.  With this choice, a {\it complete} 
 description of the evolution of the system is obtained in the shaded 
 region in the panel.  In other words, the conformal structure of the 
 {\it entire} spacetime is as in the right panel, when the system is 
 described in this reference frame.}
\label{fig:Penrose-outside}
\end{center}
\end{figure}

Let us consider a process in which a black hole is formed from a pure 
state $\ket{A}$ and then evaporates.  For simplicity, we assume that the 
black hole formed does not have a spin or charge.  We describe this process 
in a distant reference frame, i.e.\ a freely falling (local Lorentz) 
frame whose origin $p$ is outside the black hole horizon all the time; 
see the left panel of Fig.~\ref{fig:Penrose-outside}.  In its minimal 
implementation, the framework of Ref.~\cite{Nomura:2011rb} says that 
quantum states represent physical configurations on the past light cone 
of $p$ in and on the stretched/apparent horizon.%
\footnote{The stretched horizon is defined as a time-like hypersurface 
 on which the local Hawking temperature becomes of order the Planck scale 
 and thus short-distance quantum gravity effects become important (where 
 we have not discriminated between the string and Planck scales).  In 
 the Schwarzschild coordinates, it is located at $r - 2M \approx 1/M$.}
This description, therefore, represents evolution of the system in the 
shaded spacetime region in the left panel of Fig.~\ref{fig:Penrose-outside}.

An important point is that this provides a {\it complete} description 
of the {\it entire} system~\cite{'tHooft:1990fr,Susskind:1993if}---it 
is not that we describe only a part of the system corresponding to 
the shaded region; physics is complete in that spacetime region.  The 
picture describing the infalling matter inside the horizon can be obtained 
only after performing a unitary transformation on the state corresponding 
to a change of the reference frame to an infalling one~\cite{Nomura:2011rb} 
(which in general leads to a superposition of infalling and distant views, 
as will be explained in Section~\ref{sec:compl}).  In this sense, the 
{\it entire} spacetime is better represented by a Penrose diagram in the 
right panel of Fig.~\ref{fig:Penrose-outside} when the system is described 
in a distant reference frame.  As is clear from the figure, this allows 
for an $S$-matrix description of the process in Hilbert space representing 
Minkowski space ${\cal H}_{\rm Minkowski}$, which is a subspace of the 
whole covariant Hilbert space for quantum gravity: ${\cal H}_{\rm Minkowski} 
\subset {\cal H}_{\rm QG}$.  This is the case despite the fact that 
in general quantum mechanics requires only that the evolution of a 
state is unitary in the whole Hilbert space ${\cal H}_{\rm QG}$; see 
Section~\ref{sec:compl} for more discussions on this point.

What does the evolution of a quantum state look like in this description? 
Let us denote the black hole state right after the collapse of the matter 
by $\ket{{\rm BH}^0_A}$.  Since the subsequent evolution is unitary, 
the state can be written in the form $\sum_i a^t_i \ket{{\rm BH}^t_i} 
\otimes \ket{\psi^t_i}$.  Here, $\ket{{\rm BH}^t_i}$ represent states 
of the black hole (i.e.\ the horizon degrees of freedom) when time $t$ 
is passed since the formation, while $\ket{\psi^t_i}$ those of the rest 
of the world at the same time, where $t$ is the proper time measured 
at the origin of the reference frame $p$.  (The dimension of the 
Hilbert space for $\ket{{\rm BH}^t_i}$, ${\cal H}_{\rm BH}^t$, is 
$\exp({\cal A}_{\rm BH}(t)/4)$ with ${\cal A}_{\rm BH}(t) = 16\pi M(t)^2$, 
where $M(t)$ is the mass of the black hole at time $t$; the state 
$\ket{{\rm BH}^0_A}$ is an element of ${\cal H}_{\rm BH}^0$.) 
The entire state then evolves into a state representing the final 
Hawking radiation quanta, which can be written as $\sum_i a^\infty_i 
\ket{\psi^\infty_i}$.  Summarizing, the evolution of the system is 
described as
\begin{equation}
  \ket{A} \longrightarrow \ket{{\rm BH}^0_A} 
    \longrightarrow \sum_i a^t_i \ket{{\rm BH}^t_i} \otimes \ket{\psi^t_i} 
    \longrightarrow \sum_i a^\infty_i \ket{\psi^\infty_i}.
\label{eq:BH-evol}
\end{equation}
The complete information about the initial state is contained in the 
state at any time $t$ in the set of complex coefficients when the state 
is expanded in fixed basis states.  In particular, after the evaporation 
it is contained in $\{ a^\infty_i \}$ showing how the radiation states 
are superposed.

\subsection{Black hole drifting: a macroscopic uncertainty of the black 
 hole location after a long time}
\label{subsec:drifting}

What actually are the states $\ket{\psi^t_i}$?  Namely, what does the 
intermediate stage of the evaporation look like when it is described 
from a distant reference frame?  Here we argue that $\ket{\psi^t_i}$ 
for different $i$ span macroscopically different worlds.  In particular, 
the state of the black hole becomes a superposition of macroscopically 
different geometries (in the sense that they represent different 
spacetimes as viewed from the reference frame) throughout the course 
of the evaporation.  The analysis here builds upon an earlier suggestion 
by Page, who noted a large backreaction of Hawking emissions to the 
location of an evaporating black hole~\cite{Page:1979tc}.

To analyze the issue, let us take a semi-classical picture of the 
evaporation but in which the backreaction of the Hawking emission to the 
black hole energy-momentum is explicitly taken into account.  Specifically, 
we model it by a process in which the black hole emits a massless quantum 
with energy $\sim 1/M$ in a random direction in each time interval 
$\sim M$, in the rest frame of the black hole.  Here, $M$ is the mass 
of the black hole at the time of the emission.  Suppose that the velocity 
of the black hole is ${\bf v}$ before an emission; then the emission of 
a Hawking quantum will change the four-momentum of the black hole as
\begin{equation}
  p_{\rm BH}^\mu = \left( \begin{array}{cc} 
    M\gamma \\ M\gamma {\bf v} \end{array} \right) 
  \longrightarrow \left( \begin{array}{cc} 
    M\gamma - \frac{\gamma}{M}(1-{\bf n}\cdot{\bf v}) \\ 
    M\gamma {\bf v} + \frac{1}{M}{\bf n} 
      - \frac{1-\gamma}{M}\frac{{\bf n}\cdot{\bf v}}{|{\bf v}|^2} 
      {\bf v} - \frac{\gamma}{M} {\bf v} \end{array} \right),
\label{eq:BH-emit}
\end{equation}
where $\gamma \equiv 1/\sqrt{1-|{\bf v}|^2}$ and ${\bf n}$ is a unit vector 
pointing to a random direction.  The mass and the velocity of the black hole, 
therefore, change by
\begin{eqnarray}
  && \varDelta M \,=\, \sqrt{M^2-2} - M 
  \,\approx\, -\frac{1}{M},
\label{eq:delta-M}\\
  && \varDelta {\bf v} \,=\, \frac{1}{\gamma \{M^2-(1-{\bf n}\cdot{\bf v})\}}
    \left\{ {\bf n} - \left(1-\frac{1}{\gamma}\right) 
    \frac{{\bf n}\cdot{\bf v}}{|{\bf v}|^2} {\bf v} \right\}
  \,\approx\, \frac{1}{M^2}{\bf n} 
    - \frac{{\bf n}\cdot{\bf v}}{2M^2} {\bf v},
\label{eq:delta-v}
\end{eqnarray}
in each time interval
\begin{equation}
  \varDelta t \,=\, M \gamma \,\approx\, M,
\label{eq:delta-t}
\end{equation}
where we have taken the approximation that $M \gg 1$ and $|{\bf v}| \ll 1$ 
in the rightmost expressions.  In general, the emission of a Hawking 
quantum can also change the black hole angular momentum ${\bf J}$.  We 
consider this effect in the appendix, where we find that the black hole 
accumulates macroscopic angular momentum, $|{\bf J}| \gg 1$, after long 
time.  This, however, does not affect the essential part of the discussion 
below, so we will suppress it in most part.

Now, suppose that a (non-spinning) black hole is formed at $t=0$ with 
the initial mass $M_0 \equiv M(0)$.  Then, in timescales of order $M_0^3$ 
or shorter, the black hole mass is still of order $M_0$ until the very last 
moment of the evaporation.  (For example, at the Page time $t_{\rm Page} 
\sim M_0^3$, at which the black hole loses a half of its initial entropy, 
the black hole mass is still $M \approx M_0/\sqrt{2}$.)  The above process, 
therefore, can be well approximated by a process in which the black hole 
receives a velocity kick of $|\varDelta {\bf v}| \approx 1/M_0^2$ in each 
time interval $\varDelta t \approx M_0$, which after time $t$ leads to 
black hole velocity
\begin{equation}
  |{\bf v}_{\rm BH}| \approx |\varDelta {\bf v}| 
    \sqrt{\frac{t}{\varDelta t}} \sim \frac{1}{M_0^{5/2}}\sqrt{t},
\label{eq:v_BH}
\end{equation}
whose direction does not change appreciably in each kick (and so is almost 
constant throughout the process).  This implies that after time $t$ ($M_0 
\ll t \simlt M_0^3$), the location of the black hole drifts in a random 
direction by an amount
\begin{equation}
  |{\bf x}_{\rm BH}| \approx |{\bf v}_{\rm BH}| t 
    \sim \frac{1}{M_0^{5/2}} t^{3/2}.
\label{eq:x_BH}
\end{equation}
For $t \sim M_0^3$, this gives $|{\bf v}_{\rm BH}|_{t \sim M_0^3} \sim 
1/M_0$ and
\begin{equation}
  |{\bf x}_{\rm BH}|_{t \sim M_0^3} \sim M_0^2,
\label{eq:x_BH-fin}
\end{equation}
which is much larger than the Schwarzschild radius of the initial black 
hole, $R_S = 2M_0$.  By the time of the final evaporation, the velocity 
is further accelerated to $|{\bf v}_{\rm BH}| \sim 1$, but the final 
displacement is still of the order of Eq.~(\ref{eq:x_BH-fin}).

To appreciate how large the value of Eq.~(\ref{eq:x_BH-fin}) is, consider 
a black hole whose lifetime is of the order of the current age of the 
universe, $t_{\rm evap} \sim 10^{10}~{\rm years}$.  It has the initial 
mass of $M_0 \sim 10^{12}~{\rm kg}$, implying the initial Schwarzschild 
radius of $R_S \sim 1~{\rm fm}$.  The result in Eq.~(\ref{eq:x_BH-fin}) 
says that the displacement of such a black hole is $|{\bf x}_{\rm BH}| 
\sim 100~{\rm km}$ at the time of evaporation!  The origin of this 
surprisingly large effect is the longevity of the black hole lifetime, 
$t_{\rm evap} \sim M_0^3$.  For example, for a black hole of the solar 
mass $M = M_\odot \sim 10^{30}~{\rm kg}$ (i.e.\ $R_S \sim 1~{\rm km}$), 
the evaporation time is $t_{\rm evap} \sim 10^{62}~{\rm years}$---52~orders 
of magnitude longer than the age of the universe.

\begin{figure}[t]
\begin{center}
  \subfigure{\includegraphics[clip,width=.47\textwidth]{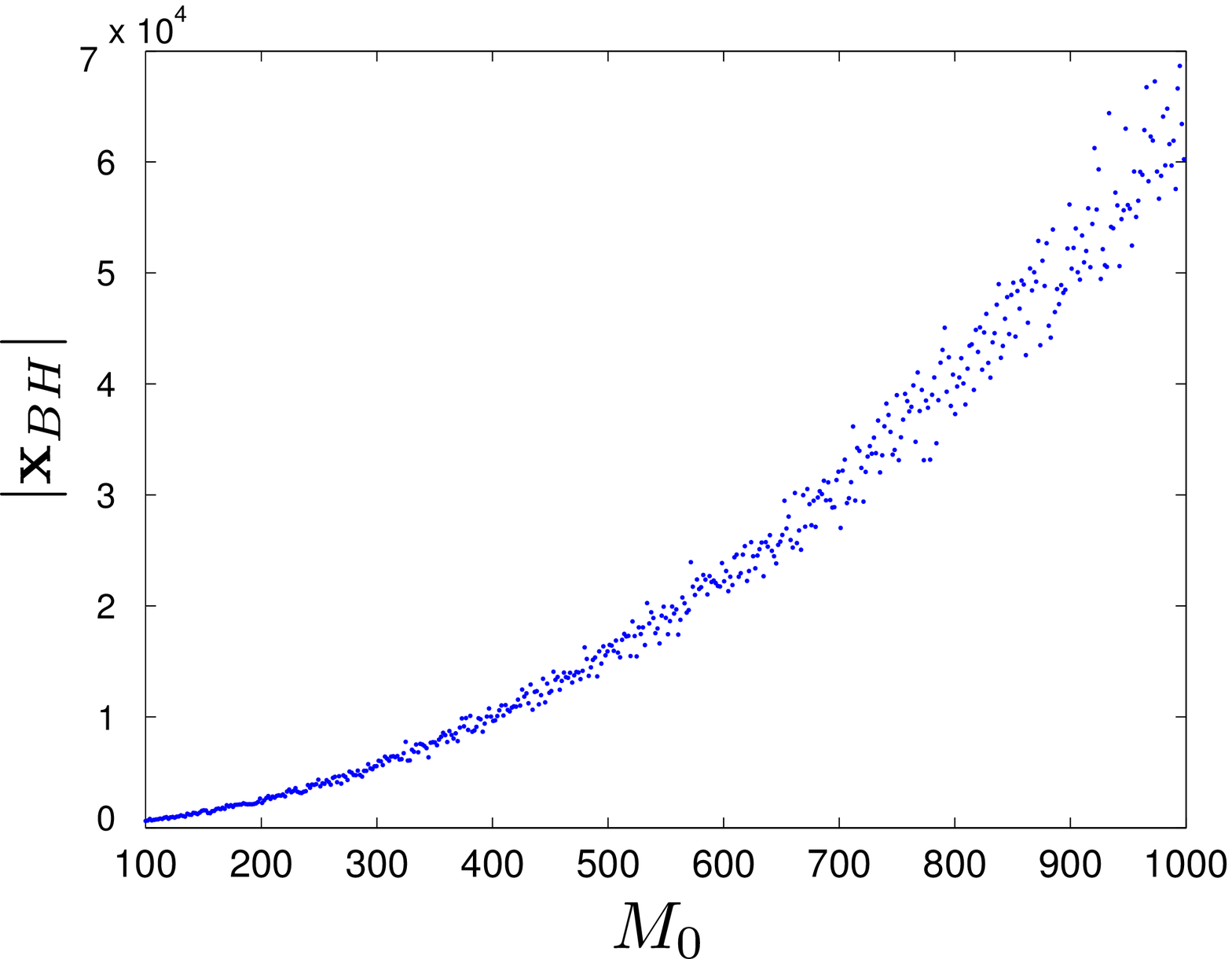}}
\hspace{1mm}
  \subfigure{\includegraphics[clip,width=.51\textwidth]{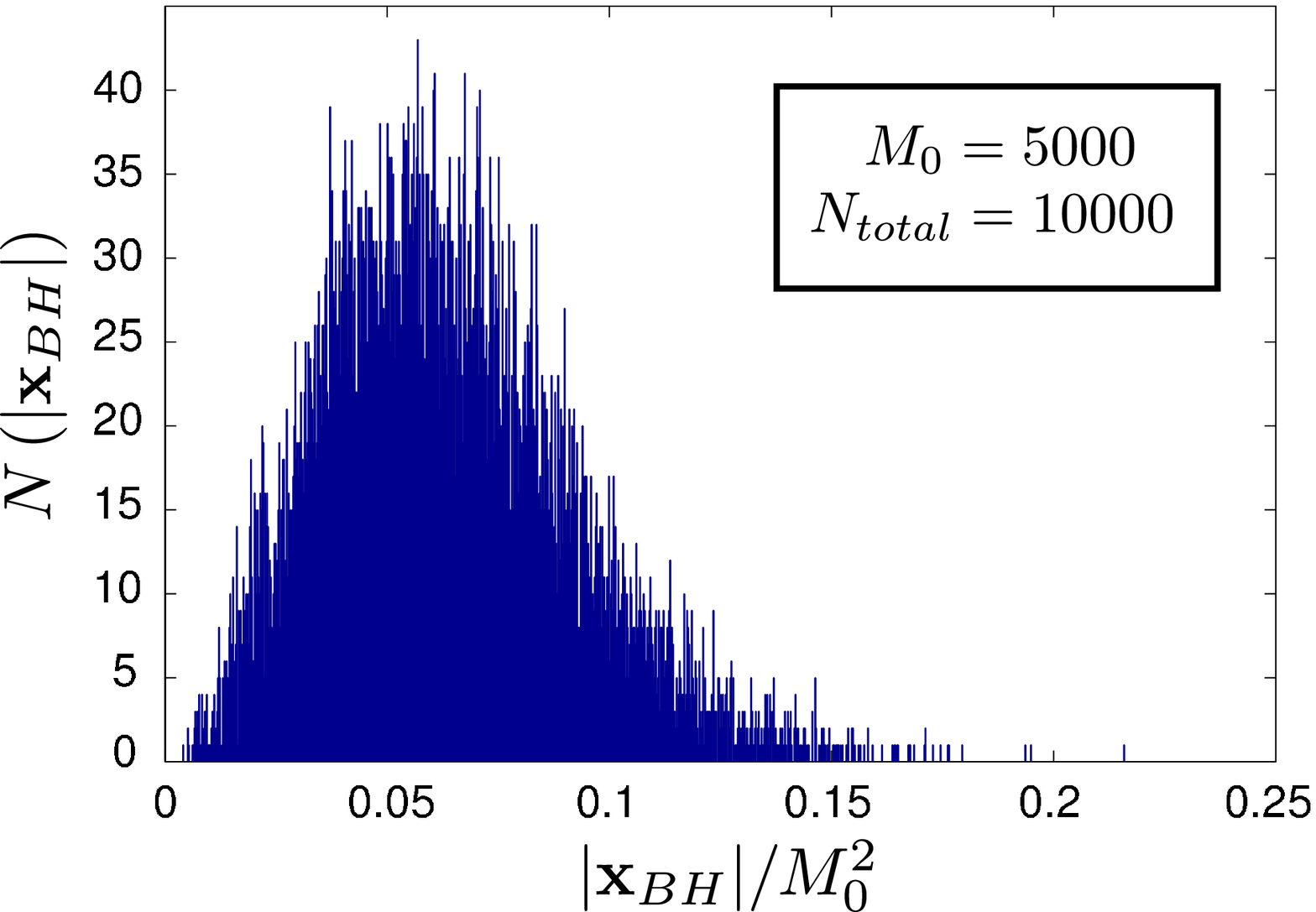}}
\caption{In the left panel, we show the result of simulating the black 
 hole displacement $|{\bf x}_{\rm BH}|$ at the Page time, $t_{\rm Page} 
 \sim M_0^3$, as a function of the initial black hole mass, $M_0$.  We 
 find the behavior expected from the general argument, $|{\bf x}_{\rm BH}| 
 \sim M_0^2$.  In the right panel, we show the probability distribution 
 of the displacement $|{\bf x}_{\rm BH}|$ for a fixed $M_0 = 5000$, obtained 
 by performing a larger number of simulations, $N_{\rm total} = 10000$. 
 The distribution takes the form expected from the central limit theorem; 
 see Eq.~(\ref{eq:x-distr}).}
\label{fig:BH-drift}
\end{center}
\end{figure}
\begin{figure}[t]
\begin{center}
  \includegraphics[width=11cm]{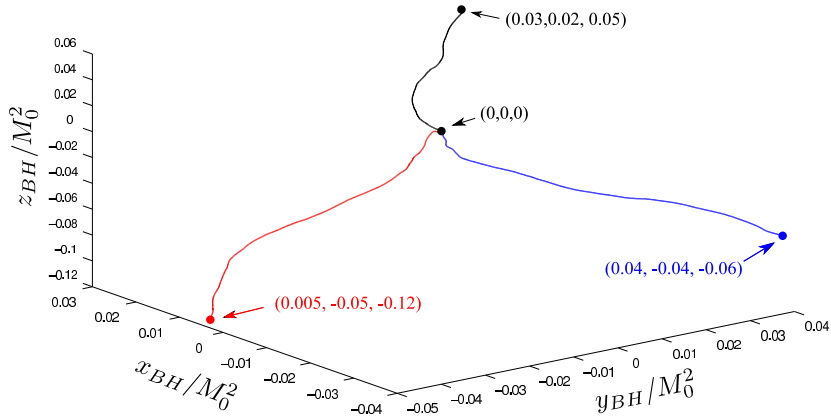}
\caption{Typical paths of the black hole drifting in the three dimensional 
 space ${\bf x}_{\rm BH} = (x_{\rm BH}, y_{\rm BH}, z_{\rm BH})$, normalized 
 by $M_0^2$.}
\label{fig:drift-3D}
\end{center}
\end{figure}

In Fig.~\ref{fig:BH-drift}, we show the result of our simulations of 
the random process described above.  In the left panel, we show the 
average value of $|{\bf x}_{\rm BH}|$ when the black hole mass is reduced 
to $M_0/\sqrt{2}$, i.e.\ at the Page time, as a function of $M_0$.  We 
see the expected behavior of $\langle |{\bf x}_{\rm BH}| \rangle \sim 
M_0^2$.  In the right panel, we show the distribution of $|{\bf x}_{\rm BH}|$ 
for a fixed $M_0$, which we take $M_0 = 5000$, with a large number of 
simulations, $N_{\rm total} = 10000$.  We find that the probability 
distribution of $|{\bf x}_{\rm BH}|$ has the form
\begin{equation}
  dP(|{\bf x}_{\rm BH}|)
  \propto |{\bf x}_{\rm BH}|^2 
    \exp\left( -c \frac{|{\bf x}_{\rm BH}|^2}{M_0^4} \right) 
    d|{\bf x}_{\rm BH}|,
\label{eq:x-distr}
\end{equation}
where $c$ is a constant of $O(1)$, as implied by the central limit theorem, 
i.e.\ each component of ${\bf x}_{\rm BH}$ having the Gaussian distribution 
centered at zero with a width $\sim M_0^2$.  We emphasize that the precise 
value of $c$ obtained by the plot does not have a physical significance, 
since it reflects our particular modeling of evaporation and omission 
of various numerical coefficients such as $8\pi$ in the expression of 
Hawking temperature $T_H = 1/8\pi M$.  Our point here is to show that 
the displacement of the black hole is indeed of $O(M_0^2)$ and its 
distribution follows what is expected from the theory of statistics. 
Note also that the reason why the left plot appears to have smaller 
distributions in $|{\bf x}_{\rm BH}|$ is because it has a smaller sample 
size; $N_{\rm total}$ for each point is of $O(10)$ in that plot.  In 
Fig.~\ref{fig:drift-3D}, we show typical paths of the black hole drift 
in three spatial dimensions.  We see that the direction of the velocity 
stays nearly constant along a path, as suggested by the general analysis.

\begin{figure}[t]
\begin{center}
  \includegraphics[width=16cm]{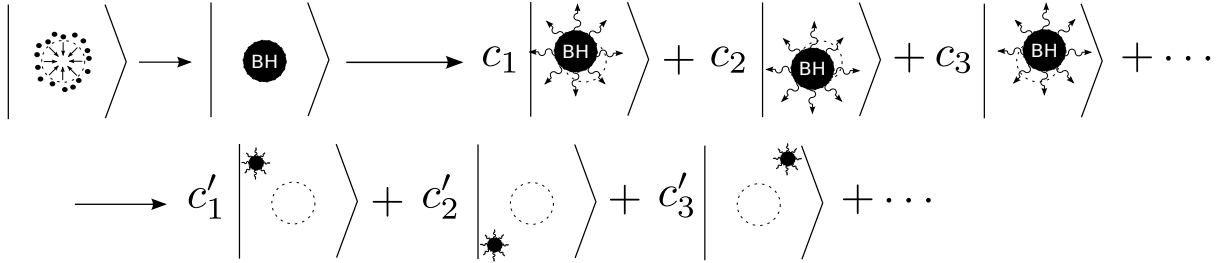}
\caption{A schematic depiction of the evolution of a black hole state 
 formed by a collapse of matter.  After long time, the state will evolve 
 into a superposition of terms representing the black hole to be in 
 macroscopically different locations, even if the initial collapsing 
 matter has a well-defined macroscopic configuration.  The variation of 
 the final locations in the evaporation timescale, $t \sim M_0^3$, is 
 of order $M_0^2$, which is much larger than the Schwarzschild radius 
 of the initial black hole, $R_S = 2M_0$.}
\label{fig:evap}
\end{center}
\end{figure}

Quantum mechanically, the result described above implies that the state 
of the black hole becomes a superposition of terms in which the black 
hole exists in macroscopically different locations, even if the initial 
state forming the black hole is a classical object having a well-defined 
macroscopic configuration.  At time $t \sim M_0^{7/3}$ after the formation 
(where $t$ is the proper time measured at $p$), the uncertainty of the 
black hole location becomes of order $M_0$, comparable to the Schwarzschild 
radius of the original black hole.  At the timescale of evaporation, 
$t \sim M_0^3$, the uncertainty is of order $M_0^2$, much larger than 
the initial Schwarzschild radius.  This is illustrated schematically 
in Fig.~\ref{fig:evap}.  Note that each term in the figure still 
represents a superposition of terms having different phase space 
configurations of emitted Hawking quanta.  Also, as shown in the 
appendix, each black hole at a fixed location is a superposition 
of black holes having macroscopically different angular momenta.

The evolution of the state depicted in Fig.~\ref{fig:evap} is obviously 
physical if we consider, for example, a super-Planckian scattering 
experiment.  In this case, we will find that Hawking quanta emitted 
at the last stage of the evaporation will come from $\sim M_0^2$ 
away from the interaction point, according to the distribution in 
Eq.~(\ref{eq:x-distr}); and we can certainly measure this because the 
wavelengths of these quanta are much smaller than $M_0^2$, and the 
interaction point is defined clearly with respect to, e.g., the beam 
pipe.  An important point here, however, is that the superposition 
nature of the black hole state is physical {\it even if there is no 
physical object other than the black hole}, e.g., the beam pipe.  This 
is because the location of an object with respect to the origin $p$ of 
the reference frame is a physically meaningful quantity in the framework 
of Ref.~\cite{Nomura:2011rb}.  In other words, the superposition nature 
discussed here is an {\it intrinsic} property of the black hole state, 
not one arising only in relation to other physical objects.

While relative values of the moduli of coefficients in front of terms 
representing different black hole locations, e.g.\ $|c_1/c_2|$ in 
Fig.~\ref{fig:evap}, are determined by the statistical analysis leading 
to Eq.~(\ref{eq:x-distr}), their relative phases are unconstrained by 
the analysis.  Moreover, it is possible that there are higher order 
corrections to the moduli that are not determined by any semi-classical 
analysis.  These quantities, therefore, can contain the information about 
the initial state; i.e., they can reflect the details of the initial 
configuration of matter that has collapsed into the black hole.  (This 
actually {\it should} be the case because a particular initial state leads 
to particular values for the relative phases because the Schr\"{o}dinger 
equation is deterministic.)  Together with the relative coefficients 
of terms representing different phase space configurations of emitted 
Hawking quanta for each black hole location (more precisely, their parts 
that are not fixed by semi-classical analyses, e.g.\ the relative phases), 
these quantities must be able to reproduce the initial state of the 
evolution by solving the appropriate Schr\"{o}dinger equation backward 
in time.

\subsection{Evolution in the covariant Hilbert space for quantum gravity}
\label{subsec:evol-H_QG}

Let us now formulate more precisely how the black hole state, formed 
by a collapse of matter, evolves in the covariant Hilbert space for 
quantum gravity, Eq.~(\ref{eq:H_QG}).  Recall that a Hilbert subspace 
${\cal H}_{\cal M}$ in Eq.~(\ref{eq:H_QG}) corresponds to the states 
realized on a fixed semi-classical three-geometry ${\cal M}$ (more precisely, 
a set of three-geometries ${\cal M} = \{ {\cal M}_i \}$ having the same 
boundary $\partial {\cal M}$).  In our context, the relevant ${\cal M}$'s 
for spacetime with the black hole are specified by the location of the 
black hole ${\bf x}_{\rm BH}$ (which can be parameterized, e.g., by the 
direction $\{ \theta, \phi \}$ and the affine length $\lambda$ of the 
past-directed light ray connecting reference point $p$ to the closest 
point on the stretched horizon) and the size of the black hole (which 
can be parameterized, e.g., by its mass $M$ or area ${\cal A} = 16\pi 
M^2$).  Here and below, we ignore the angular momentum of the black 
hole, for simplicity.  We also need to consider the Hilbert subspace 
corresponding spacetime without the black hole, ${\cal H}_0$.

The part of ${\cal H}_{\rm QG}$ relevant to our problem here is then
\begin{equation}
  {\cal H} = \left( \bigoplus_{{\bf x}_{\rm BH},\, M} 
    {\cal H}_{{\bf x}_{\rm BH},\, M} \right) \oplus {\cal H}_0,
\label{eq:H}
\end{equation}
where $0 < M \leq M_0$, and we have used the notation in which 
${\bf x}_{\rm BH}$ and $M$ are discretized.  The Hilbert subspace 
${\cal H}_{{\bf x}_{\rm BH},\, M}$ consists of the factor associated 
with the black hole horizon ${\cal H}_{{\bf x}_{\rm BH},\, M}^{\rm 
horizon}$ and that with the rest ${\cal H}_{{\bf x}_{\rm BH},\, 
M}^{\rm bulk}$ (which represents the region outside the horizon):
\begin{equation}
  {\cal H}_{{\bf x}_{\rm BH},\, M} = 
    {\cal H}_{{\bf x}_{\rm BH},\, M}^{\rm horizon} \otimes
    {\cal H}_{{\bf x}_{\rm BH},\, M}^{\rm bulk}.
\label{eq:H-xM}
\end{equation}
According to the Bekenstein-Hawking entropy, the size of the horizon factor 
is given by
\begin{equation}
  {\rm dim}\, {\cal H}_{{\bf x}_{\rm BH},\, M}^{\rm horizon} 
    = e^{\frac{{\cal A}_{\rm BH}}{4}}
    = e^{4\pi M^2},
\label{eq:dim-horizon}
\end{equation}
regardless of ${\bf x}_{\rm BH}$.  Because of this, Hilbert space 
factors ${\cal H}_{{\bf x}_{\rm BH},\, M}^{\rm horizon}$ for different 
${\bf x}_{\rm BH}$ are all isomorphic with each other, which allows us 
to view ${\cal H}_{{\bf x}_{\rm BH},\, M}^{\rm horizon}$ for any fixed 
${\bf x}_{\rm BH}$ as the intrinsic structure of the black hole.

Now, right after the formation of the black hole, which we assume to 
have happened at ${\bf x}_{{\rm BH},0}$ at $t=0$, the system is in a state 
that is an element of ${\cal H}_{{\bf x}_{{\rm BH},0},\, M_0}$.  In the 
case of Eq.~(\ref{eq:BH-evol})
\begin{equation}
  \ket{\Psi(0)} \equiv \ket{{\rm BH}^0_A} 
    \in {\cal H}_{{\bf x}_{{\rm BH},0},\, M_0}.
\label{eq:Psi-0}
\end{equation}
This state then evolves into a superposition of states in different 
${\cal H}_{\cal M}$'s.%
\footnote{This is precisely analogous to the case of $e^+e^-$ scattering, 
 in which the initial state $\ket{e^+e^-} \in {\cal H}_2$ evolves into 
 a superposition of states in different ${\cal H}_n$'s, e.g. $\ket{e^+e^-} 
 \rightarrow c_e \ket{e^+e^-} + \cdots + c_{ee} \ket{e^+e^-e^+e^-} + 
 \cdots$, where ${\cal H}_n$ is the $n$-particle subspace of the entire 
 Fock space: ${\cal H} = \bigoplus_n {\cal H}_n$.}
At time $t$, the state of the system can be written as
\begin{equation}
  \ket{\Psi(t)} = \sum_{{\bf x}_{\rm BH}} \alpha^t_{{\bf x}_{\rm BH}} 
    \ket{\phi^t_{{\bf x}_{\rm BH}}},
\label{eq:Psi-t}
\end{equation}
where $\ket{\phi^t_{{\bf x}_{\rm BH}}} \in {\cal H}_{{\bf x}_{\rm BH},\, 
M(t)}$, and we have ignored possible fluctuations of the black hole 
mass at a fixed time $t$, for simplicity. (Including this effect 
is straightforward; we simply have to add terms corresponding to 
${\cal H}_{{\bf x}_{\rm BH},\, M}$ with $M \neq M(t)$.)  The state 
$\ket{\phi^t_{{\bf x}_{\rm BH}}}$ contains the horizon {\it and} other 
degrees of freedom, according to Eq.~(\ref{eq:H-xM}).  We can expand 
it in some basis in ${\cal H}_{{\bf x}_{\rm BH},\, M(t)}^{\rm horizon}$ 
(e.g.\ the one spanned by states having well-defined numbers of Hawking 
quanta emitted afterward) or in some basis in ${\cal H}_{{\bf x}_{\rm 
BH},\, M(t)}^{\rm bulk}$ (e.g.\ the one spanned by states having 
well-defined phase space configurations of already emitted Hawking 
quanta).  In either case, it takes the form
\begin{equation}
  \ket{\phi^t_{{\bf x}_{\rm BH}}} = \sum_n \beta^t_n 
    \ket{{\rm BH}^t_{{\bf x}_{\rm BH},n}} \otimes 
    \ket{\psi^t_{{\bf x}_{\rm BH},n}},
\label{eq:phi}
\end{equation}
where $\ket{{\rm BH}^t_{{\bf x}_{\rm BH},n}} \in {\cal H}_{{\bf x}_{\rm 
BH},\, M(t)}^{\rm horizon}$ and $\ket{\psi^t_{{\bf x}_{\rm BH},n}} \in 
{\cal H}_{{\bf x}_{\rm BH},\, M(t)}^{\rm bulk}$.  Plugging this into 
Eq.~(\ref{eq:Psi-t}) and defining
\begin{equation}
  a^t_i \equiv \alpha^t_{{\bf x}_{\rm BH}} \beta^t_n,
\label{eq:ati}
\end{equation}
where $i \equiv \{ {\bf x}_{\rm BH}, n \}$, we reproduce the third 
expression in Eq.~(\ref{eq:BH-evol}).  In this formulation, the statement 
that the black hole state is a superposition of macroscopically different 
geometries refers to the fact that coefficients $|\alpha^t_{{\bf x}_{\rm 
BH}}|$ have a significant support in a wide range of ${\bf x}_{\rm BH}$ 
extending beyond the original Schwarzschild radius $M_0$.

\subsection{What does a physical observer actually see?}
\label{subsec:what_see}

We have found that a late black hole state is far from a
semi-classical state in which the spacetime has a fixed geometry;
rather, it involves a superposition of macroscopically different
geometries.  Does this mean that a physical observer sees something
very different from what the usual picture based on general relativity
predicts?

The answer is no.  To understand this, let us consider a physical observer 
watching the evaporation process from a distance by measuring (all or parts 
of) the emitted Hawking quanta.  For simplicity, we consider that he/she 
does that using usual measuring devices, e.g.\ by locating photomultipliers 
around the black hole from which he/she collects the data.  This leads 
to an entanglement between the system and the observer (or his/her brain 
states).  And because the interactions leading to it are local, the observer 
is entangled with the basis in ${\cal H}_{{\bf x}_{\rm BH},\, M}^{\rm bulk}$ 
spanned by the states that have well-defined phase space configurations 
of emitted Hawking quanta (within the errors dictated by the uncertainty 
principle) and well-defined locations for the black hole (since the 
black hole location can be inferred from the momenta of the Hawking 
quanta)~\cite{Nomura:2011rb}.  Namely, the combined state of the black 
hole and the observer evolves as
\begin{equation}
  \ket{{\rm BH}^0_A} \otimes \bigl| \observer \bigr> 
  \longrightarrow
  \sum_{{\bf x}_{\rm BH},\, n} a^t_{{\bf x}_{\rm BH},n} 
    \ket{{\rm BH}^t_{{\bf x}_{\rm BH},n}} \otimes 
    \ket{\psi^t_{{\bf x}_{\rm BH},n}} \otimes
    \bigl| \observer\brain \bigr>,
\label{eq:obs-percep}
\end{equation}
where $ \ket{\psi^t_{{\bf x}_{\rm BH},n}}$ represents the state in which 
the black hole is in a well-defined location ${\bf x}_{\rm BH}$ and Hawking 
quanta have a well-defined phase space configuration $n$.  The last factor 
in the right-hand side implies that the observer recognized that the black 
hole is at ${\bf x}_{\rm BH}$ and the configuration of emitted Hawking 
quanta is $n$.

Since terms in the right-hand side of Eq.~(\ref{eq:obs-percep}) have 
macroscopically different configurations, e.g.\ the brain state of the 
observer differs, their mutual overlaps are exponentially suppressed 
(e.g.\ by $\sim \prod_{i=1}^N \epsilon_i$, where $\epsilon_i < 1$ is 
the overlap of each atom and $N$ the total number of atoms).  The observer 
in each term (or branch), therefore, sees his/her own universe; i.e., 
the interferences between different terms are negligible.  For any 
of these observers, the behavior of the black hole is controlled by 
semi-classical physics (but with the backreaction of the emission taken 
into account).  For instance, they all see that the black hole keeps 
emitting Hawking quanta consistent with the thermal spectrum with 
temperature $T_H(t) = 1/8\pi M(t)$, and that it drifts in a fixed direction 
as a result of backreactions, eventually evaporating at a location 
$\sim M_0^2$ away from that of the formation.  A single observer cannot 
predict the direction to which the black hole will drift, reflecting 
the fact that the entire state is a superposition of terms having 
different $({\bf x}_{\rm BH} - {\bf x}_{{\rm BH},0})/|({\bf x}_{\rm BH} 
- {\bf x}_{{\rm BH},0})|$, but all these observers find a set of common 
properties for the black hole, including the relation between $T_H$ and 
$M$.%
\footnote{More precisely, there are rare observers who find deviations 
 from these relations, but the probability for that to happen is 
 exponentially suppressed.}

It is these ``intrinsic properties'' of the black hole that the 
semi-classical gravity on a fixed Schwarzschild geometry (in which 
the black hole is located at the ``center'') really describes.  A physical 
observer watching the evolution does not see anything contradicting 
what is implied by the semi-classical analysis about these intrinsic 
properties.  This is true despite the fact that the full quantum state 
obtained by evolving collapsing matter that initially had a well-defined 
configuration takes the form in Eqs.~(\ref{eq:Psi-t},~\ref{eq:phi}), 
which involves a superposition of macroscopically different geometries 
and is very different at late times from a ``semi-classical state'' 
having a fixed geometry.

\subsection{Can a physical observer recover the information?}
\label{subsec:info_recov}

The black hole evaporation process is often compared with burning a book 
in classical physics:\ if we measure all the details of the emitted Hawking 
quanta, we can recover the initial state from these data by solving the 
Schr\"{o}dinger equation backward in time.  Is this correct?

It is true that if we know the coefficients of all the terms in a state 
when it is expanded in a fixed basis, e.g.\ $\{ a^\infty_i \}$ in 
Eq.~(\ref{eq:BH-evol}), then unitarity must allow us to recover the 
initial state unambiguously.  However, a physical observer measuring 
Hawking radiation from black hole evaporation can {\it never} obtain 
the complete information about these coefficients, even if he/she measures 
{\it all} the radiation quanta.  In the state in Eq.~(\ref{eq:obs-percep}), 
for example, a physical observer ``lives'' in {\it one} of the terms in 
the right-hand side and, therefore, cannot have the information about 
the coefficients of the other terms.  The other terms are already 
decohered---or ``decoupled''---so that they are other worlds/universes 
for the observer.

In fact, the situation is exactly the same in usual scattering experiments. 
Consider two initial states $\ket{e^+e^-}$ and $\ket{\mu^+\mu^-}$ with the 
same $\sqrt{s}$ ($> 2 m_\tau$) and angular momentum.  They evolve as
\begin{eqnarray}
  \ket{e^+e^-} &\longrightarrow& a_1 \ket{e^+e^-} + a_2 \ket{\mu^+\mu^-} 
    + a_3 \ket{\tau^+\tau^-} + \cdots,
\\
  \ket{\mu^+\mu^-} &\longrightarrow& b_1 \ket{e^+e^-} + b_2 \ket{\mu^+\mu^-} 
    + b_3 \ket{\tau^+\tau^-} + \cdots,
\end{eqnarray}
where we have ignored the momenta and spins of the final particles. 
The information about an initial state is in the {\it complete} set 
of coefficients in the final superposition state; i.e., if we know 
the entire $\{ a_i \}$ (or $\{ b_i \}$), then we can recover the initial 
state by solving the evolution equation backward.  However, if a physical 
observer measures a final state, e.g., as $\tau^+\tau^-$, how can he/she 
know that it has arisen from $e^+e^-$ or $\mu^+\mu^-$ scattering?  In 
general, if an observer measures the final outcome of a process, he/she 
will be entangled with one of the terms in the final state (in the above 
case, $\ket{\tau^+\tau^-}$), so there is no way that he/she can learn 
all the coefficients in the final state.

The situation does not change even if the observer uses a carefully-crafted 
quantum device which, upon interacting with the radiation, is entangled 
{\it not} with a well-defined phase space configuration of the radiation 
quanta but with a macroscopic superposition of those configurations.  In 
this case, the basis of the final state to which the observer is entangled 
may be changed, but it still cannot change the fact that he/she will be 
entangled with {\it one} of the terms in the final state, i.e., he/she 
will measure {\it a} possible outcome among all the possibilities.

Therefore, in quantum mechanics, an observer can never recover the 
initial state by observing the final state.  The statement that the 
final state of an evolution contains all the information about the initial 
state is {\it not} the same as the statement that a physical observer 
measuring the final state can recover the initial state if he/she measures 
a system with high enough (or even infinite) precision.  The only way 
that an observer can test the relation between the initial and final states 
is to {\it create} the same initial state many times and perform multiple 
(including quantum) measurements on the final states.  (Note that creating 
many initial states in this context differs from producing a copy of 
a generic unknown state, which is prohibited by the quantum no-cloning 
theorem~\cite{Wootters:1982zz}.)  A single system does not allow for 
doing this, no matter how high the precision of the measurement is, 
and no matter how clever the measurement device is.

\section{Complementarity as a Reference Frame Change}
\label{sec:compl}

So far, we have been describing the formation and evaporation of a 
black hole from a distant reference frame.  In this reference frame, 
the complete description of the process is obtained in the spacetime 
region outside and on the (time-like) stretched horizon, where 
intrinsically quantum gravitational---presumably stringy (such as 
fuzzball~\cite{Lunin:2001jy})---effects become important.  What then 
is the significance of the interior of the black hole horizon, where 
we expect to have regular low-curvature spacetime according to general 
relativity?

As discussed in Ref.~\cite{Nomura:2011rb}, and implied by the original 
complementarity picture~\cite{Susskind:1993if}, a description of the 
internal spacetime is obtained (only) after changing the reference frame. 
An important point is that the reference frame change is represented 
as a unitary transformation acting on a quantum state, so if we want 
to discuss the precise mapping between the pictures based on different 
reference frames, then we need to keep all the terms in the state. 
In this section, we carefully study issues associated with the reference 
frame changes, especially in describing an old black hole.

\subsection{Describing the black hole interior}
\label{subsec:interior}

Suppose collapsing matter, which initially had a well-defined classical 
configuration, forms a black hole, which then eventually evaporates. 
In a distant reference frame, this process is described as in 
Eq.~(\ref{eq:BH-evol}), which we denote by $\ket{\Psi(t)}$.  How 
does the process look from a different reference frame?

Since a reference frame can be any freely falling (locally Lorentz) frame, 
the new description can be obtained by performing a translation, rotation, 
or boost on a quantum state at fixed $t$~\cite{Nomura:2011rb}.  In general, 
the state on which these transformations act, however, contains the 
horizon degrees of freedom as well as the bulk ones.  How do they 
transform under the transformations?

We do not know the microscopic description of the horizon degrees of 
freedom or their explicit transformations under the reference frame 
changes.  Nevertheless, we can know which spacetime regions are transformed 
to which horizon degrees of freedom, and vice versa, by {\it assuming} 
that the global spacetime picture in semi-classical gravity is consistent 
with the one obtained by a succession of these reference frame changes. 
Here we phrase this in the form of a hypothesis:
\begin{itemize}
\item[] {\bf Complementarity Hypothesis:}
 The transformation laws of a quantum state under the reference frame 
 changes are consistent with those obtained in the global spacetime picture 
 based on general relativity.  In particular, the transformation laws 
 between the horizon and bulk degrees of freedom are constrained by this 
 requirement.
\end{itemize}
As discussed in Refs.~\cite{Nomura:2011rb,Nomura:2011dt}, this hypothesis 
is fully consistent with the holographic principle formulated in the form 
of the covariant entropy conjecture~\cite{Bousso:1999xy}.  Specifically, 
the dimension of the Hilbert space representing horizon degrees of freedom 
and that representing the corresponding spacetime region before (or 
after) a transformation are the same for general spacetimes, including 
the cosmological ones, as it should be.  Alternatively, we can take a 
view that if we require that the above hypothesis is true in the covariant 
Hilbert space ${\cal H}_{\rm QG}$, then the covariant entropy conjecture 
is obtained as a consequence.

Let us now consider a reference frame change induced by a boost performed 
at some early time $t_{\rm boost} < 0$ (before the black hole forms at 
$t=0$) in such a way that the origin $p$ of the reference frame enters 
the black hole horizon at some late time $t_{\rm enter} > 0$.  In this 
subsection, we focus on the case
\begin{equation}
  t_{\rm enter} \ll M_0^{7/3},
\label{eq:early-ent}
\end{equation}
so that the uncertainty of the black hole location at the time when $p$ 
enters the horizon is negligible, and we ignore the (exponentially) small 
probability that $p$ misses the horizon.  (The possibility of $p$ missing 
the horizon becomes important when we discuss the description of the 
interior of an older black hole.)

\begin{figure}[t]
\begin{center}
  \includegraphics[width=12cm]{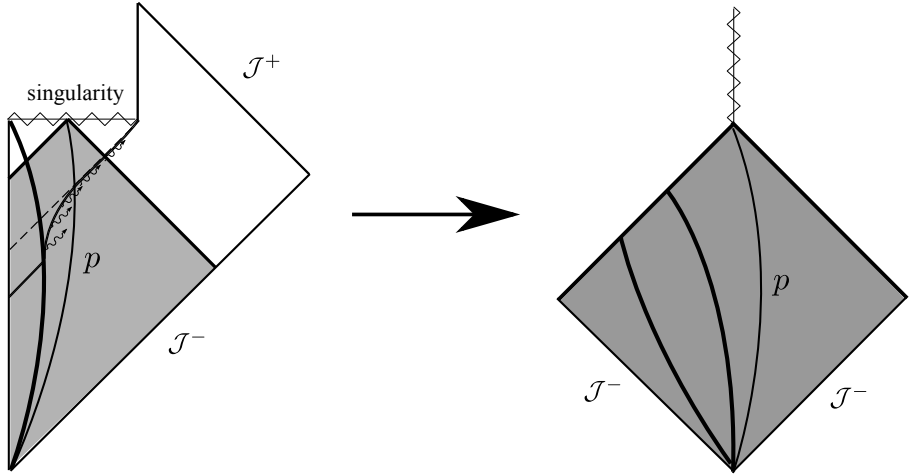}
\caption{The left panel shows the standard global spacetime picture 
 for the formation and evaporation of a black hole, with the shaded 
 region representing the spacetime region described by an infalling 
 reference frame.  (The trajectory of the origin, $p$, of the reference 
 frame is also depicted.)  As discussed in the text, this is the 
 {\it entire} spacetime when the system is described in this reference 
 frame, so its conformal structure is in fact as in the right panel. 
 Here, the wavy line with a solid core represents singularity states.}
\label{fig:Penrose-inside}
\end{center}
\end{figure}

Recall that quantum states in the present framework represent physical 
configurations on the past light cone of $p$ (in and on the apparent 
horizon) when they allow for spacetime interpretation, i.e.\ when the 
curvature at $p$ is smaller than the Planck scale.  Therefore, the spacetime 
region represented by the state of the system after the reference frame 
change
\begin{equation}
  \ket{\Psi'(t)} = e^{-iH(t-t_{\rm boost})} U_{\rm boost} 
    e^{iH(t-t_{\rm boost})} \ket{\Psi(t)},
\label{eq:ref-change}
\end{equation}
where $U_{\rm boost}$ is the boost operator represented in 
${\cal H}_{\rm QG}$, corresponds to the shaded region in the left 
panel of Fig.~\ref{fig:Penrose-inside}.
Specifically, $\ket{\Psi'(t)}$ at
\begin{equation}
  t < t_{\rm enter} + t_{\rm fall}
\label{eq:spacetime}
\end{equation}
describes this region, with $t_{\rm fall} \approx O(M_0)$ being 
the time needed for $p$ to reach the singularity after it passes the 
horizon.  After $t = t_{\rm enter} + t_{\rm fall}$, the state evolves 
in the Hilbert subspace ${\cal H}_{\rm sing}$, which consists of 
states that are associated with spacetime singularities and thus do 
not allow for spacetime interpretation.  The detailed properties of 
these ``intrinsically quantum gravitational'' states are unknown, 
except that ${\rm dim}\, {\cal H}_{\rm sing} = \infty$, implying that 
generic singularity states do not evolve back to the usual spacetime 
states~\cite{Nomura:2011rb}.

In the right panel of Fig.~\ref{fig:Penrose-inside}, we depict the 
causal structure of the spacetime as viewed from the new reference frame. 
Because of the lack of the spherical symmetry, we have depicted the region 
swept by two past-directed light rays emitted from $p$ in the opposite 
directions (while in the left panel we have depicted only the region with 
fixed angular variables with respect to the center of mass of the system). 
The singularity states are represented by the wavy line with a solid core 
at the top.  Note that, as in the case of the description in the distant 
reference frame (depicted in Fig.~\ref{fig:Penrose-outside}), this is 
the {\it entire} spacetime region when the system is described in this 
infalling reference frame---the non-shaded region in the left panel 
simply does not exist.  (Including the non-shaded region, indeed, is 
overcounting as indicated by the standard argument of information cloning 
in black hole physics.)  A part of the non-shaded region appears if we 
change the reference frame, but only at the cost of losing some of the 
shaded region.  The global spacetime picture in the left panel appears 
only if we ``patch'' the views from different reference frames, which, 
however, grossly overcounts the correct quantum degrees of freedom.

There are two comments.  First, the reference frame change considered 
here is (obviously) only {\it a} reference frame change among possible 
(continuously many) reference frame changes, all of which lead to 
different descriptions of the same physical process.  Second, a unitary 
transformation representing this reference frame change,
\begin{equation}
  U(t) = e^{-iH(t-t_{\rm boost})} U_{\rm boost} e^{iH(t-t_{\rm boost})},
\label{eq:U-ref}
\end{equation}
does not close in the Hilbert space ${\cal H}$ in Eq.~(\ref{eq:H}), 
although it closes in the whole covariant Hilbert space ${\cal H}_{\rm QG}$. 
Before the reference frame change, the evolution of the state is 
given by a trajectory in ${\cal H} = ( \oplus_{{\bf x}_{\rm BH},\, M} 
{\cal H}_{{\bf x}_{\rm BH},\, M} ) \oplus {\cal H}_0$.  The action 
of $U(t)$ maps this into a trajectory in
\begin{equation}
  {\cal H}' = {\cal H}_0 \oplus {\cal H}_{\rm sing},
\label{eq:H'}
\end{equation}
with $\ket{\Psi'(-\infty)} \in {\cal H}_0$ and $\ket{\Psi'(+\infty)} 
\in {\cal H}_{\rm sing}$.%
\footnote{Note that ${\cal H}_0$ contains a set of states that represent 
 three-geometries whose {\it boundary} (at an infinity) is that of the 
 flat space, i.e. a two-dimensional section of ${\cal J}^-$.}
As a result, in this new reference frame, the evolution of the 
system does not allow for an $S$-matrix description in ${\cal H}_0$ 
(or ${\cal H}_{\rm Minkowski}$), although it still allows for an 
``$S$-matrix'' description in the whole ${\cal H}_{\rm QG}$ (or in 
${\cal H}_{\rm Minkowski} \oplus {\cal H}_{\rm sing}$), which 
contains the singularity states in ${\cal H}_{\rm sing}$.

\subsection{Complementarity for an old black hole}
\label{subsec:compl-old}

Let us now try to describe the interior of an older black hole, specifically 
the spacetime inside the black hole horizon after a time $> O(M_0^{7/3})$ 
is passed since the formation.  To do this, we can consider performing 
a boost at time $t_{\rm boost} < 0$ on $\ket{\Psi(t)}$ in such a way that 
$p$ enters the black hole horizon at time $t_{\rm enter} \gg M_0^{7/3}$. 
What does the resultant state $\ket{\Psi'(t)}$ look like?

As discussed in the previous subsection, this can be done by applying 
an operator of the form of Eq.~(\ref{eq:U-ref}) on $\ket{\Psi(t)}$, where 
$U_{\rm boost}$ now represents a different amount of boost than the one 
considered before.  In general, the relation between the states before 
and after a reference frame change is highly nontrivial.  For example, 
time $t$ is measured by the proper time at $p$, but relations between 
the proper times of the two frames depend on the geometries as well as 
the paths of $p$ therein.  Therefore, various terms in $\ket{\Psi'(t)}$ 
for a fixed $t$ may correspond to terms in $\ket{\Psi(t)}$ of different 
$t$'s.  Without knowing the explicit form of $H$ and $U_{\rm boost}$ 
represented in the whole ${\cal H}_{\rm QG}$, which includes the horizon 
degrees of freedom, how can we know the form of the state after the 
transformation?

According to our complementarity hypothesis, the probability of finding 
a certain history for the evolution of geometry must agree in the two 
pictures before and after the reference frame change if the geometries 
are appropriately transformed, i.e.\ according to the global spacetime 
picture in general relativity.  To elucidate this, let us consider the 
black hole evolution described in Eqs.~(\ref{eq:Psi-t},~\ref{eq:phi}) in 
a distant reference frame, and ask what is the probability that the black 
hole follows a particular path ${\bf r}(t)$ in a time interval between 
$t_I$ and $t_F$ within the error $|\varDelta {\bf r}| < \epsilon(t)$. 
For simplicity, we do this by requiring that the black hole satisfies the 
above conditions at discretized times $t_i$; $i=0,\cdots,N$ ($\gg 1$), 
with $t_0 \equiv t_I$ and $t_N \equiv t_F$.  The probability is then 
given by
\begin{equation}
  P = \prod_{i=0}^{N}
    \left( \sum_{|\varDelta {\bf r}| < \epsilon_i} 
      |\alpha^{t_i}_{{\bf r}_i + \varDelta{\bf r}}|^2 \right),
\label{eq:BH-traj-1}
\end{equation}
where ${\bf r}_i \equiv {\bf r}(t_i)$ and $\epsilon_i \equiv \epsilon(t_i)$. 
This provides the probability of a particular semi-classical history to 
appear, given the state $\ket{\Psi(t)}$.  We can now ask a similar question 
for the state $\ket{\Psi'(t)}$:\ what is the probability of having the 
black hole to follow the trajectory ${\bf r'}(t)$ between $t'_I$ and $t'_F$ 
within the error $\epsilon'(t)$?  The resulting probability is
\begin{equation}
  P' = \prod_{i=0}^{N}
    \left( \sum_{|\varDelta {\bf r'}| < \epsilon'_i} 
      |\alpha^{t_i}_{{\bf r'}_i + \varDelta{\bf r'}}|^2 \right),
\label{eq:BH-traj-2}
\end{equation}
where $t_0 = t'_I$ and $t_N = t'_F$.  The complementarity hypothesis in 
the previous subsection asserts that the two probabilities are the same
\begin{equation}
  P = P',
\label{eq:P=P'}
\end{equation}
if the relation between $\{ {\bf r}(t), \epsilon(t), t_I, t_F \}$ and 
$\{ {\bf r'}(t), \epsilon'(t), t'_I, t'_F \}$ is the one obtained by 
performing the corresponding transformation in general relativity on 
the semi-classical background selected by Eq.~(\ref{eq:BH-traj-1}).

\begin{figure}[t]
\begin{center}
  \includegraphics[width=16cm]{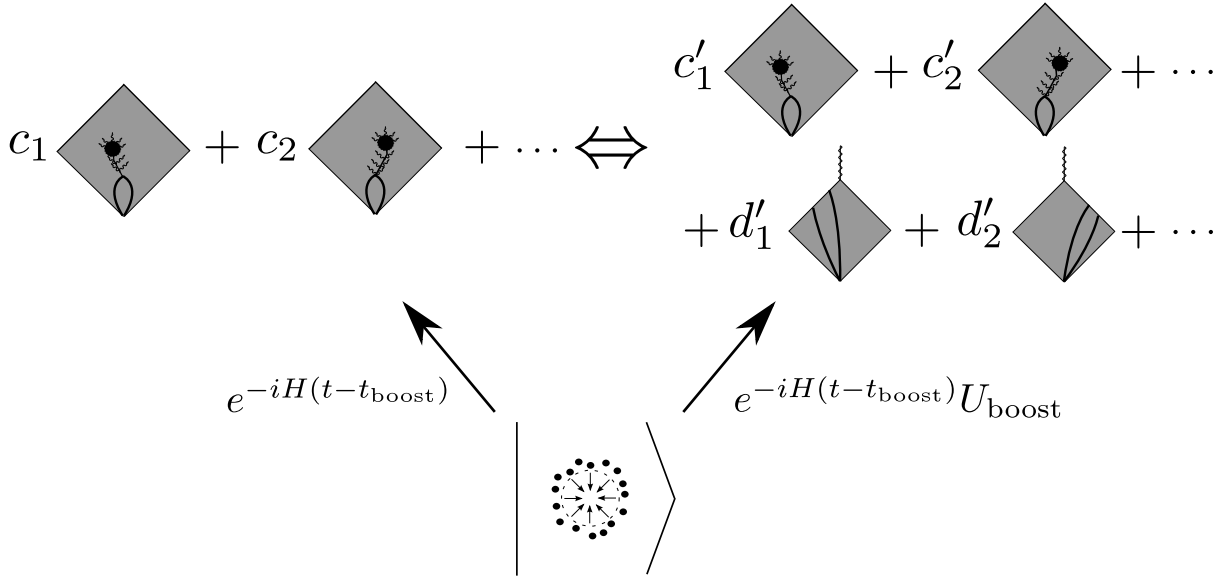}
\caption{A schematic picture of a relation between the two descriptions 
 based on two different reference frames of an old black hole formed by 
 collapsing matter that initially had a well-defined classical configuration. 
 In one reference frame, the black hole is viewed from outside, and the 
 state becomes a superposition of black holes in different locations 
 at late times (depicted schematically in the left-hand side).  In the 
 other reference frame, obtained by acting $U_{\rm boost}$ on the state 
 at $t_{\rm boost}$, the reference point $p$ enters the black hole horizon 
 at late time $t_{\rm enter} \gg M_0^{7/3}$, allowing for a description 
 of internal spacetime (in the right-hand side).  This, however, happens 
 only for some of the terms, depicted in the second line, since $p$ misses 
 the horizon in most of the terms because of the large uncertainty of 
 the black hole location, i.e.\ $\sum_i |d'_i|^2 \ll \sum_i |c'_i|^2$.}
\label{fig:compl-1}
\end{center}
\end{figure}

The above analysis implies that when we perform a boost on $\ket{\Psi(t)}$ 
at an early time $t_{\rm boost} < 0$, trying to describe the interior of 
an old black hole with $t_{\rm enter} \gg M_0^{7/3}$, then the resultant 
state can only be a {\it superposition} of infalling and distant descriptions 
of the process, since in most of the semi-classical histories represented 
by $\ket{\Psi(t)}$, the trajectory of $p$ obtained by the boost will miss 
the black hole horizon because of the large uncertainty of the black hole 
location.  Namely, complementarity obtained by this reference frame change 
is the one between the distant description and the superposition of the 
infalling and distant descriptions specified by the state $\ket{\Psi'(t)}$. 
This is illustrated schematically in Fig.~\ref{fig:compl-1}.

\begin{figure}[t]
\begin{center}
  \includegraphics[width=14cm]{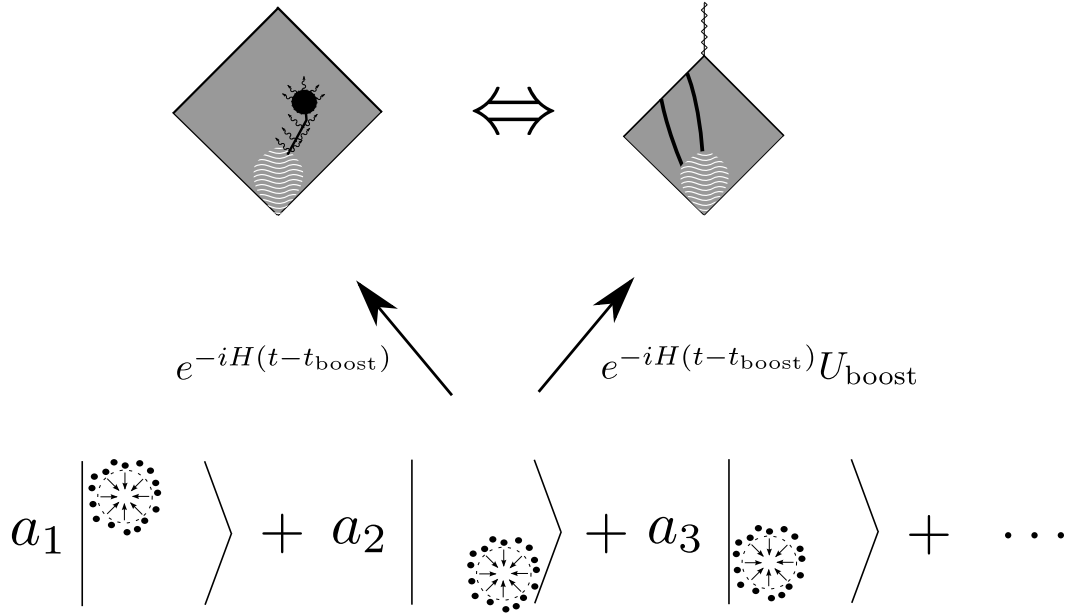}
\caption{A complementarity relation between the internal and external 
 descriptions of an old black hole can be obtained if we consider a state 
 in which the black hole has a well-defined semi-classical configuration 
 at late time $t_{\rm enter}$.  Such a state, however, can arise through 
 evolution only if we consider a special initial state in which coefficients 
 $a_i$ of the terms representing well-defined configurations of collapsing 
 matter are finely-adjusted so that the state represents a black hole in 
 a well-defined location at late time $\sim t_{\rm enter}$.  (The regions 
 with wavy white lines indicate superpositions of classical geometries.)}
\label{fig:compl-2}
\end{center}
\end{figure}

Is it possible to obtain a direct correspondence between the interior and 
exterior of an old black hole, without involving a superposition?  This 
can be done if we focus only on a term in $\ket{\Psi(t)}$ in which $p$ 
just misses the black hole horizon, with the smallest distance between 
$p$ and the horizon achieved at some time $t_{\rm min} \gg M_0^{7/3}$. 
We can then evolve this term slightly backward in time, to $t_{\rm boost} 
= t_{\rm min} - \epsilon$ ($\epsilon \ll M_0^{7/3}$), and perform a boost 
there so that $p$ enters into the horizon at some time after $t_{\rm boost}$. 
In this way, the correspondence between the terms representing the interior 
and exterior can be obtained.  An important point, however, is that neither 
of these terms can be obtained by evolving initial collapsing matter 
that had a well-defined classical configuration (which would lead to 
a superposition of the black hole in vastly different locations).  Rather, 
by evolving the state further back beyond $t_{\rm boost}$, we would obtain 
a superposition of states each of which represents collapsing matter 
with a well-defined classical configuration.  (This state would have 
finely-adjusted coefficients so that after evolving to $t_{\rm boost} 
\sim t_{\rm min}$, the black hole is in a well-defined location with 
respect to $p$.)  This situation is illustrated in Fig.~\ref{fig:compl-2}.

The discussion above implies that there is no well-defined complementarity 
map between the interior and exterior of an old black hole throughout the 
course of the black hole evolution within the purely semi-classical picture. 
Such a map must involve a superposition of semi-classical geometries at 
some point in the evolution.  We note that while the state in the intermediate 
stage of the evolution can be a superposition of elements in ${\cal H}_0$, 
${\cal H}_{{\bf x}_{\rm BH},\, M}$, and ${\cal H}_{\rm sing}$, it becomes 
a superposition of elements in ${\cal H}_0$ and ${\cal H}_{\rm sing}$ 
at $t \rightarrow \infty$.  Therefore, the ``$S$-matrix'' description 
discussed in the previous subsection is still available in this case in 
the Hilbert space of ${\cal H}_{\rm Minkowski} \oplus {\cal H}_{\rm sing}$.

\subsection{On firewalls (or firewall as an exponentially unlikely phenomenon)}
\label{subsec:firewall}

The complementarity picture has recently been challenged by the ``firewall 
paradox'' posed by Almheiri, Marolf, Polchinski, and Sully (AMPS) 
in Ref.~\cite{Almheiri:2012rt}.  Here we elucidate the discussion of 
Ref.~\cite{Nomura:2012sw}, refuting one of the AMPS arguments based 
on measurements of early Hawking radiation.  AMPS also has another 
argument using entropy relations, building on an earlier discussion by 
Mathur~\cite{Mathur:2009hf,Braunstein:2009my}.  Here we focus only on 
the first AMPS argument based on measurements.  Addressing the second 
one requires an additional assumption about the decoherence structure 
of microscopic degrees of freedom of the horizon, beyond what we have 
postulated in this paper~\cite{Nomura:2012ex}.

In essence, the argument by AMPS goes as follows.  Consider an old black 
hole with $t > t_{\rm Page}$ that has formed from collapse of some pure 
state.  Because of the purity of the state, the system as viewed from 
a distant reference frame can be written as
\begin{equation}
  \ket{\Psi} = \sum_i c_i \ket{i} \otimes \ket{\psi_i},
\label{eq:state}
\end{equation}
where $\ket{i} \in {\cal H}_{\rm horizon}$ and $\ket{\psi_i} \in 
{\cal H}_{\rm rad}$ represent degrees of freedom associated with the 
horizon region and the emitted Hawking quanta.  (For simplicity we have 
suppressed the time index, which is not essential for the discussion 
here.)  For a black hole older than $t_{\rm Page}$, the dimensions of 
the Hilbert space factors satisfy ${\rm dim}\,{\cal H}_{\rm horizon} 
\ll {\rm dim}\,{\cal H}_{\rm rad}$.  Therefore, states $\ket{\psi_i}$ 
for different $i$ are expected to be nearly orthogonal, and one can 
construct a projection operator $P_i$ that acts only on ${\cal H}_{\rm 
rad}$ (not on ${\cal H}_{\rm horizon}$) but selects a term in 
Eq.~(\ref{eq:state}) corresponding to a specific state $\ket{i}$ 
in ${\cal H}_{\rm horizon}$ when operated on $\ket{\Psi}$:
\begin{equation}
  P_i \ket{\Psi} \propto \ket{i} \otimes \ket{\psi_i}.
\label{eq:projection}
\end{equation}
The point is that one can construct such an operator for an arbitrary 
state $\ket{i}$ in ${\cal H}_{\rm horizon}$.

AMPS argue that since the infalling observer can access the early 
radiation, he/she can select a particular term in Eq.~(\ref{eq:state}) 
by making a measurement on those degrees of freedom.  In particular, they 
imagine that such a measurement would select a term in which $\ket{i}$ 
in ${\cal H}_{\rm horizon}$ is an eigenstate, $\ket{\tilde{\imath}}$, of 
the number operator, $b^\dagger b$, of a Hawking radiation mode that will 
escape from the horizon region:
\begin{equation}
  b^\dagger b \ket{\tilde{\imath}} \propto \ket{\tilde{\imath}}.
\label{eq:i-tilde}
\end{equation}
If this were true, then the infalling observer must find physics 
represented by $\ket{\tilde{\imath}}$ near the horizon, and since 
an eigenstate of $b^\dagger b$ cannot be a vacuum for the infalling 
modes $a_\omega$, related to $b$ by
\begin{equation}
  b = \int_0^\infty\! d\omega 
    \left( B(\omega) a_\omega + C(\omega) a_\omega^\dagger \right)
\label{eq:b_vs_a}
\end{equation}
with some functions $B(\omega)$ and $C(\omega)$, the infalling observer 
must experience nontrivial physics at the horizon (i.e.\ $a_\omega 
\ket{\tilde{\imath}} \neq 0$ for infalling modes with the frequencies 
much larger than the inverse horizon size).  This obviously contradicts 
what is expected from general relativity.

As discussed in Ref.~\cite{Nomura:2012sw}, this argument misses the 
fact that the emergence of a classical world in the underlying quantum 
world is {\it a dynamical process} dictated by unitary evolution of 
a state, and {\it not} something we can impose from outside by acting 
with some projection operator on the state.  In particular, the existence 
of the projection operator $P_i$ for an arbitrary $i$ does not imply 
that a measurement performed by a {\it classical} observer, which general 
relativity is supposed to describe, can pick up the corresponding state 
$\ket{i}$.  To understand this point, consider a state representing 
a superposition of upward and downward chairs (relative to some other 
object, e.g.\ the ground, which we omit):
\begin{equation}
  \ket{\Psi_{\rm chair}} = \bigl| \chairup \bigr> + \bigl| \chairdown \bigr>.
\label{eq:chair}
\end{equation}
An observer interacting with this system evolves following the unitary, 
deterministic Schr\"{o}dinger equation; in particular, the combined 
chair and observer state becomes
\begin{equation}
  \ket{\Psi_{\rm chair + observer}} 
  = \bigl| \chairup \bigr> \otimes \bigl| \observer\brainup \bigr> 
    + \bigl| \chairdown \bigr> \otimes \bigl| \observer\braindown \bigr>.
\label{eq:chair+obs}
\end{equation}
This does {\it not} lead to a classical world in which the chair is in 
a superposition state but to two different worlds in which the chair 
is upward and downward, respectively.  Namely, the measurement is performed 
in the particular basis $\bigl\{ \bigl| \chairup \bigr>, \bigl| \chairdown 
\bigr> \bigr\}$, which is {\it determined by the dynamics}---the existence 
of an operator projecting onto a superposition chair state does not 
mean that a measurement is performed in that basis.  For sufficiently 
macroscopic object/observer, the appropriate basis for measurements 
is almost always (see below) the one in which they have well-defined 
configurations in classical phase space (within some errors, which 
must exist because of the uncertainty principle).  This is because the 
Hamiltonian has the form that is local in spacetime~\cite{Nomura:2011rb}.

In the specific context of the firewall argument, a measurement of 
the early radiation by an infalling, classical observer will select 
a state in ${\cal H}_{\rm rad}$ that has a well-defined classical 
configuration of Hawking radiation quanta $\ket{\psi_I}$, because 
interactions between him/her and the quanta are local.  As shown in 
Ref.~\cite{Nomura:2012sw}, the state $\ket{\psi_I}$ selected in this 
way is expected {\it not} to be a state that is maximally entangled 
with $\ket{\tilde{\imath}}$, i.e.\ $\ket{\psi_{\tilde{\imath}}}$.  In 
other words, the basis in ${\cal H}_{\rm rad}$ selected by entanglement 
with the eigenstates of $b^\dagger b$ is different from the one selected 
by entanglement with the infalling classical observer.  This implies 
that in the world described by the infalling observer, i.e.\ in a term 
of the entire state in which the observer has a well-defined classical 
configuration, the state in ${\cal H}_{\rm rad}$ is always in a 
superposition of $\ket{\psi_{\tilde{\imath}}}$'s for different 
$\tilde{\imath}$'s, so that the corresponding state in ${\cal H}_{\rm 
horizon}$ is {\it not} an eigenstate of $b^\dagger b$.  In particular, 
there is no contradiction if the state is a simultaneous eigenstate 
of $a_\omega$'s with the eigenvalue zero (up to exponentially small 
corrections) as implied by general relativity.%
\footnote{Of course, we are only showing here that the argument of AMPS 
 breaks down, i.e.\ the complementarity picture---or the semi-classical 
 picture in the infalling reference frame---is {\it consistent}. 
 Proving it, e.g.\ showing that the state is indeed a simultaneous 
 eigenstate of $a_\omega$'s, would require an understanding of the 
 underlying theory of quantum gravity.  As stated explicitly in 
 Section~\ref{subsec:interior}, complementarity is a hypothesis, 
 which we argue is a consistent one~\cite{Nomura:2012sw}.}

One might ask what happens if we prepare a carefully-crafted {\it quantum} 
device that will be entangled with one of the $\ket{\psi_{\tilde{\imath}}}$'s 
and then send a signal, e.g.\ a particle, toward the horizon.  Wouldn't 
that particle see a firewall at the horizon?  Yes, that particle 
might see a firewall, but it is not a phenomenon described by general 
relativity, a theory for a classical world.  First of all, in order for 
the device to be entangled with $\ket{\psi_{\tilde{\imath}}}$, it must 
collect the information in the Hawking quanta to learn that they are in 
the $\ket{\psi_{\tilde{\imath}}}$ state, and since the information is 
encoded in a highly scrambled form, it must be very large collecting 
many quanta spread in space without losing their coherence.  This 
implies that the device must have been in an extremely carefully-chosen 
superposition state of different classical configurations at the beginning 
of the evolution (which is also clear from the fact that ${\rm dim}\,{\cal 
H}_{\rm horizon} \ll {\rm dim}\,{\cal H}_{\rm rad} \simlt {\rm dim}\,{\cal 
H}_{\rm device}$, where ${\cal H}_{\rm device}$ is the Hilbert space factor 
associated with the device).  Now, we know that if the initial state is 
extremely fine-tuned, an extremely unlikely event can happen.  For example, 
if the initial locations and velocities of the molecules are finely tuned, 
ink dissolved in a water tank can spontaneously come to a corner.  The 
firewall phenomenon is analogous to this kind of phenomena.

More specifically, we can ask what is the degree of fine-tuning needed 
to see firewalls.  The amount of fine-tuning, i.e.\ the probability for 
a randomly-chosen device state to see the firewall, is estimated as
\begin{equation}
  p_{\rm fw} \sim {\rm dim}\,{\cal H}_{\rm horizon} 
    \sqrt{{\rm dim}\,{\cal H}_{\rm rad}}\,
    \epsilon^{{\rm dim}\,{\cal H}_{\rm rad}},
\qquad
  \epsilon < 1,
\label{eq:firewall-prob}
\end{equation}
which is obtained by asking the probability of a randomly-chosen 
basis state for measuring Hawking radiation to agree with one of the 
$\ket{\psi_{\tilde{\imath}}}$'s; see the appendix of Ref.~\cite{Nomura:2012sw} 
for a similar calculation.  Since ${\rm dim}\,{\cal H}_{\rm rad} \simgt 
e^{2\pi M_0^2}$ after the Page time, this is double-exponentially 
suppressed in a macroscopic number $M_0^2 \gg 1$:
\begin{equation}
  p_{\rm fw} \simlt \epsilon^{e^{2\pi M_0^2}} \lll 1,
\label{eq:no-firewall}
\end{equation}
analogous to the case of having an entropy decreasing process in 
a system with large degrees of freedom.

Note that if we ask the amount of fine-tuning needed for ink dissolved 
in a water tank to come to a corner in the context of classical physics, 
we would find it to be suppressed single-exponentially by a large number, 
$\sim \epsilon^{N_A}$ where $N_A$ is Avogadro's number.  This is because 
states having classically well-defined configurations are already 
exponentially rare in the whole Hilbert space, although they are {\it 
dynamically selected}---if we instead ask the fraction of the whole 
quantum states (including arbitrary superposition states) leading to ink 
in the corner, we would obtain a number with double-exponential suppression 
as in Eq.~(\ref{eq:no-firewall}).  In the context of the firewall, the 
initial device state is intrinsically quantum mechanical, so we {\it must} 
fine-tune the initial condition at the level of Eq.~(\ref{eq:no-firewall}), 
i.e.\ we cannot use dynamical selection to reduce the fine-tuning.  In 
this sense, the amount of fine-tuning needed to see the firewall phenomenon 
is even worse than that needed to see an entropy decreasing phenomenon 
in usual classical systems.

\section{Discussion and Conclusions}
\label{sec:concl}

In this paper, we have described a complete evolution---the formation 
and evaporation---of a black hole in the framework of quantum gravity 
preserving locality, given in Ref.~\cite{Nomura:2011rb}.  While some 
of the results obtained are indeed specific to this context, some are 
more general applying to other theories of gravity as well, especially 
to the ones in which the formation and evaporation of a black hole is 
described as a unitary quantum mechanical process.  Key ingredients to 
understand these results are
\begin{itemize}
\item[(i)]
The system must be described in a fixed reference frame.  Moreover, the 
complete physical description of a process is obtained in the spacetime 
region in and on the stretched/apparent horizon {\it as viewed from 
that reference frame}.  In particular, in the minimal implementation 
of the framework of Ref.~\cite{Nomura:2011rb}, quantum states correspond 
to physical configurations on the past light cone of a fixed reference 
point $p$ in and on the horizon.
\item[(ii)]
A quantum state is in general a superposition of terms representing 
macroscopically different configurations/geometries.  This is true 
despite the fact that the dynamics, represented by the time evolution 
operator, is local.  In fact, interactions between degrees of freedom 
generically lead to such a superposition because of a distinct feature 
of quantum mechanics:\ rapid amplification of information to a more 
macroscopic level.  (Measurements are a special case of this more 
general phenomenon.)
\end{itemize}

The element in (i) implies that the global spacetime picture of general 
relativity is an ``illusion'' in the sense that the internal and future 
asymptotically-flat spacetimes do not exist simultaneously.  In one 
description based on a distant reference frame, there is only external 
spacetime and the process can be described by a unitary $S$-matrix in 
${\cal H}_{\rm Minkowski}$, as depicted in Fig.~\ref{fig:Penrose-outside}. 
In another description based on an infalling frame, the interior spacetime 
does exist but there is no future asymptotic Minkowski region, so 
the process should be described in the whole covariant Hilbert space 
${\cal H}_{\rm QG}$ containing both ${\cal H}_{\rm Minkowski}$ and 
${\cal H}_{\rm sing}$; see Fig.~\ref{fig:Penrose-inside}.  The global 
spacetime in general relativity provides a picture in which these and 
other {\it equivalent} descriptions are {\it all} represented at the 
same time, so it grossly overcounts the true quantum degrees of freedom.

A virtue of the general relativistic description, however, lies in the 
fact that it correctly represents observations made by a {\it classical} 
observer traveling in spacetime.  In particular, the equivalence 
principle correctly captures the fact that the infalling classical 
observer does not see anything unusual at the horizon.  An important 
point here is that the classical observer---or more generally a 
classical world---{\it emerges} through unitary evolution of a 
full quantum state as a basis state in which the information is 
amplified~\cite{q-Darwinism,Nomura:2011rb}.  Observations made by 
any observer, detector, etc.\ in such a world are well described by 
general relativity.  While a phenomenon contradicting a generic prediction 
of general relativity could in principle occur~\cite{Almheiri:2012rt}, 
such a situation requires an exponentially fine-tuned initial 
condition~\cite{Nomura:2012sw}, analogous to an (exponentially 
unlikely) entropy decreasing process in a system with large degrees 
of freedom.

The element in (ii) is important at least for two reasons.  First, 
this provides an additional place in which the information about 
the initial state is encoded in a black hole final state, in particular 
in relative phases of the coefficients of the terms representing 
macroscopically different black hole configurations.  Indeed, a late 
black hole state becomes a superposition of black holes with different 
spins and in different locations, even if the back hole is formed 
from collapsing matter that had a well-defined classical configuration 
without an angular momentum.  Second, this aspect of the evolution also 
affects the way in which a complementarity map between interior and 
exterior spacetimes works for an old black hole.  Because of the branching 
of the black hole state described above, such a map must involve a 
superposition of semi-classical geometries at some point in the evolution. 
Namely, there is no simple correspondence between the two semi-classical 
regions/geometries for an old black hole that is applicable throughout 
the entire history of the evolution (even if we focus on time before 
the reference point hits the singularity).

In treating a system with gravity, it is often assumed that there is a 
fixed semi-classical background geometry.  For example, this is the case 
in the original complementarity argument~\cite{Susskind:1993if} and in 
most treatments of defining probabilities in the eternally inflating 
multiverse~\cite{Guth:2000ka}.  At first sight, this does not lead to 
any problem because the relevant systems are large---we usually do not 
need to keep track of the whole quantum nature of the state in those 
cases, including a superposition of possible outcomes, entanglement 
with the observer, and so on.  These intrinsically quantum properties, 
however, are crucial when we discuss the {\it fundamental structure} 
of the theory such as unitarity and information.  As discussed 
in Section~\ref{subsec:info_recov}, by focusing on a particular 
outcome---which is a completely legitimate procedure in discussing 
the outcome of a particular experiment---we will never be able to 
see the correct unitary structure of the underlying quantum theory. 
Committing to a specific semi-classical geometry is precisely such 
a treatment.  An important message is that avoiding these treatments, 
i.e.\ keeping the full superposition---or {\it many worlds}---of the 
state, is a key to evade many apparent problems/paradoxes in black hole 
physics~\cite{Nomura:2012sw} and in eternally-inflating multiverse 
cosmology~\cite{Nomura:2011dt,Nomura:2011rb}.  Hopefully, the present 
paper adds further clarifications on this issue, and provides a useful 
framework for further studies of fundamental issues in black hole 
physics.

\section*{Acknowledgments}

This work was supported in part by the Director, Office of Science, Office 
of High Energy and Nuclear Physics, of the US Department of Energy under 
Contracts DE-FG02-05ER41360 and DE-AC02-05CH11231, by the National Science 
Foundation under grants PHY-0855653 and DGE-1106400, and by the Simons 
Foundation grant 230224.

\appendix

\section{Spontaneous Spin-up of a Schwarzschild Black Hole}
\label{app:spin}

Just as a black hole accumulates momentum over its lifetime through 
randomly recoiling from Hawking emissions, we can ask if a black hole 
also accumulates angular momentum due to the spin and orbital angular 
momentum of emitted particles.  In this appendix, we argue that the 
answer is yes:\ non-rotating black holes with initial mass $M_0$ 
spontaneously spin up to angular momentum $J \equiv |{\bf J}| \sim 
M_0$ at a time of order $M_0^3$.  This implies that a Schwarzschild black 
hole evolves into a superposition of Kerr black holes with different 
values of ${\bf J}$, although the resulting angular momenta will be 
small enough, $J/M^2 \ll 1$, that the geometry of each term is still 
well approximated by the Schwarzschild one.

To begin with, let us consider how many Hawking quanta are emitted by 
the time at which an initial black hole loses some fixed fraction of 
its mass, e.g.\ the Page time at which the black hole mass becomes 
$M = M_0/\sqrt{2}$.  The number of emitted quanta is
\begin{equation}
  N \sim \frac{M_0}{T_H} \sim M_0^2,
\label{eq:app-N}
\end{equation}
where $T_H \sim 1/M_0$ is the Hawking temperature.  If the emitted 
quanta consist of a particle with spin $s > 0$, then each emission 
changes the angular momentum of the black hole by $\varDelta J \sim s$, 
depending on the direction of the spin.  Assuming that the emission is 
unbiased in the direction of angular momentum (see below), we find that 
the black hole accumulates the angular momentum
\begin{equation}
  J \sim s \sqrt{N} \sim M_0,
\label{eq:app-J}
\end{equation}
at a time of order $M_0^3$, where we have taken $s \sim O(1)$ in the 
last expression.

If the Hawking quanta consist of a scalar ($s=0$), then most of the 
emissions do not affect the black hole angular momentum since the 
emissions are dominated by $s$-wave.  However, there is a small 
probability that a quantum is emitted in a higher angular momentum 
mode.  The probability is dominated by $p$-wave ($l=1$), which can 
be calculated for small $J/M^2$ as $p \simeq 0.002 + O(J/M^2)$, 
independent of $M$~\cite{Starobinskii:1973,Page:1976df}.  Therefore, 
the number of Hawking quanta that affect the black hole angular 
momentum is $p N$, and the accumulated angular momentum of the 
black hole is
\begin{equation}
  J \sim \sqrt{p N} \sim M_0,
\label{eq:app-J-scalar}
\end{equation}
which is parametrically the same as in the case of a particle with spin.

One might think that once the accumulated angular momentum becomes 
macroscopic, $J \gg 1$, the black hole becomes a Kerr black hole, so 
that there is a bias in the Hawking spectrum that preferentially selects 
emissions that reduce $J$~\cite{Hawking:1974sw}, preventing a further 
accumulation of $J$.  We now argue, however, that until the time $t \sim 
M_0^3$ when the mass of the black hole starts decreasing significantly, 
the evolution of ${\bf J}$ is well approximated by a random walk process 
as described above.

\begin{figure}[t]
\begin{center}
  \includegraphics[width=8cm]{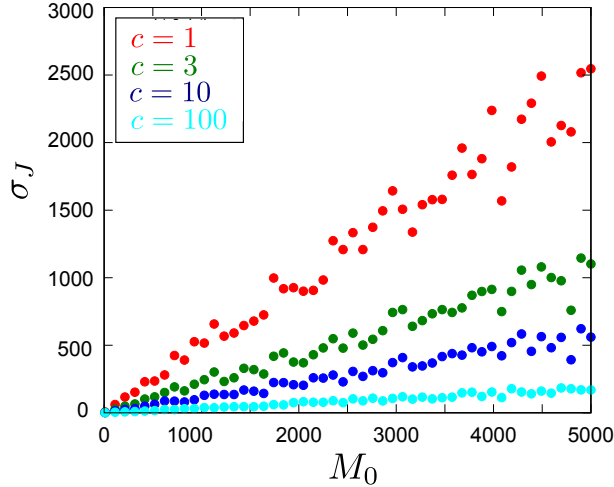}
\caption{Plot of $\sigma_J$, the square root of the variance of $J$ at 
 the Page time, as a function of the initial mass $M_0$.  Each data point 
 represents $\sigma_J$ obtained by $100$ simulations.  The different colors 
 correspond to different values of coefficient $c$ in Eq.~(\ref{eq:J-bias}), 
 which measures the strength of the angular momentum emission bias from 
 a Kerr black.}
\label{fig:angdat}
\end{center}
\end{figure}

To see this, at a given time $t$, let us call the direction of ${\bf J}$ 
the $z$-axis.  Suppose an emission of a particle with spin $s$ changes 
$J = J_z$, which occurs with $O(p)$ and $O(1)$ probabilities for $s = 0$ 
and $s > 0$, respectively.  For small $J/M^2$, the probability $\rho_+$ 
($\rho_{-}$) that the emission increases (reduces) $J$ is~\cite{Page:1976df}:
\begin{equation}
  \rho_{\pm} = \frac{1}{2} \mp c \frac{J}{M^2},
\label{eq:J-bias}
\end{equation}
where $J$ and $M$ are the magnitude of angular momentum and the mass 
before the emission takes place, and $c$ is an $O(1)$ coefficient 
which depends on the type of a particle emitted and is independent of 
$J$ and $M$ to first order in $J/M^2$. Numerical simulations of this 
process indicate that this bias is not strong enough to prevent a black 
hole from spinning up to $J \sim M_0$ by the Page time, $t_{\rm Page}$. 
Results of these simulations are shown in Fig.~\ref{fig:angdat}, where 
we have assumed a change of $J$ according to Eq.~(\ref{eq:J-bias}) in 
each time interval $M_0$.  The results indicate that
\begin{equation}
  J \sim f(c) M_0 \sim M_0
\label{eq:final-J}
\end{equation}
at $t \sim t_{\rm Page}$, where $f$ is a monotonically decreasing function 
of $c$; in fact, our simulations suggest that $f(c) \propto 1/\sqrt{c}$ 
for $c \simgt 1$.

The results obtained above can be understood by the following simple 
argument.  Imagine that at some late time $t \simgt M_0^3$, the probability 
distribution of the black hole angular momentum reaches some ``equilibrium'' 
distribution $P(J)$, in which the random walk effect increasing $J$ 
is balanced with the bias of the emission reducing $J$.  According to 
Eq.~(\ref{eq:J-bias}), this implies
\begin{equation}
  \rho_+ P(J) = \rho_- P(J+1),
\label{eq:balance}
\end{equation}
leading to
\begin{equation}
  \frac{P(J+1)}{P(J)} = \frac{1-2c\frac{J}{M^2}}{1+2c\frac{J+1}{M^2}} 
  \approx 1 - 4c \frac{J}{M^2}.
\label{eq:balance-2}
\end{equation}
Here, we have used $1 \ll J \ll M^2$ in the last expression.  This has 
the solution
\begin{equation}
  P(J) \sim e^{-2c \frac{J^2}{M^2}}.
\label{eq:balance-sol}
\end{equation}
Namely, the black hole angular momentum has a characteristic size
\begin{equation}
  J \sim \frac{1}{\sqrt{c}} M,
\label{eq:J}
\end{equation}
consistent with the result obtained in Eq.~(\ref{eq:final-J}).

In summary, we conclude that a Schwarzschild black hole with initial mass 
$M_0$ will spontaneously spin up to $J \sim M_0$ by a timescale of order 
$M_0^3$.  When the black hole mass starts decreasing significantly, its 
angular momentum will also start decreasing, following Eq.~(\ref{eq:J}). 
The combination $J/M^2$ keeps increasing as $1/M$ but is still (much) 
smaller than $1$, as long as $M \gg 1$ where our analysis is valid. 
What happens at the real end of the evaporation is unclear, but we can 
say that while the evolution of a Schwarzschild black hole leads to a 
superposition of Kerr black holes with distinct angular momenta, the 
probability of it becoming a macroscopic extremal black hole ($J = M^2 
\gg 1$) is, most likely, exponentially suppressed.

\end{document}